\documentclass[a4paper,11pt]{article}

\usepackage{amsmath}
\usepackage{amsthm}
\usepackage{makeidx}
\usepackage{graphicx}
\usepackage{amsfonts}
\usepackage{amssymb}
\usepackage{fancyhdr}
\usepackage{epsfig}
\usepackage{slashbox}
\usepackage[nottoc]{tocbibind}
\usepackage{hyperref}
\usepackage[rightcaption]{sidecap}
\usepackage[usenames,dvipsnames]{color}
\usepackage{authblk}

\newcommand{\nexp}[1]{\,\mbox{\rm e}^{\displaystyle{#1}}}
\newcommand{\iu}{\mbox{\rm i\,}}

\newcommand{\labfig}[1]{\label{#1fig}}
\newcommand{\reffig}[1]{Figure~\protect\ref{#1fig}}

\newcommand{\FigSideHeight}[3]{
\begin{SCfigure}[50][!Hhtb]
\centering \epsfig{file=#1.eps,height=#2} \hspace{0.5cm}
\caption{\footnotesize #3} \labfig{#1}
\end{SCfigure}
}

\newcommand{\FigSideWidth}[3]{
\begin{SCfigure}[50][!Hhtb]
\centering \epsfig{file=#1.eps,width=#2} \hspace{0.5cm}
\caption{\footnotesize #3} \labfig{#1}
\end{SCfigure}
}

\newcommand{\FIGh}[3]{
\begin{figure}[!Hhbt]
\vspace{2.0mm}
\begin{center}
\epsfig{file=#1.eps,width=#2,clip}
\end{center}
\vspace{-0.7cm} \caption{\footnotesize #3} \labfig{#1}
\end{figure}
}

\begin{document}

\title{A variational mode solver for optical waveguides based on
quasi-analytical vectorial slab mode expansion}
\author[1]{Alyona Ivanova}
\author[1,2]{Remco Stoffer}
\author[1]{Manfred Hammer}

\affil[1]{MESA+ Institute for Nanotechnology, University of
Twente, Enschede, The Netherlands}

\affil[2]{PhoeniX Software, Enschede, The Netherlands}

\date{}
\maketitle

\graphicspath{{Pictures/}{Pictures/SquareWaveguide/}{Pictures/SquareWaveguide/Fields_30modes/}{Pictures/Covergences/Box/}{Pictures/Covergences/Slanted/}{Pictures/Covergences/Diffused/}{Pictures/diffused_fields/}{Pictures/RibWaveguide/RibWaveguideConvergence/}{Pictures/PolConv/}}

\noindent{\bfseries A flexible and efficient method for fully
vectorial modal analysis of 3D dielectric optical waveguides with
arbitrary 2D cross-sections is proposed. The technique is based on
expansion of each modal component in some a priori defined
functions defined on one coordinate axis times some unknown
coefficient-functions, defined on the other axis. By applying a
variational restriction procedure the unknown
coefficient-functions are determined, resulting in an optimum
approximation of the true vectorial mode profile. This technique
can be related to both Effective Index and Mode Matching methods.
A couple of examples illustrate the performance of the method.}

\section{Introduction}
\label{sec:intro}

Three-dimensional optical channel waveguides are basic components
of integrated optical devices such as directional couplers,
wavelength filters, phase shifters, and optical switches. The
successful design of these devices requires an accurate estimation
of the modal field profiles and propagation constants. Over
already some decades several classes of methods for the analysis
of dielectric optical waveguides were developed: among these are
techniques of more numerical character, like Finite Element and
Finite Difference approximations, the Method of Lines, and
Integral Equations Methods, but also more analytical approaches
like Film Mode Matching (FMM) and the Effective Index Method
(EIM). Detailed overviews of these techniques can be found in
\cite{Chi94,Vas97,SGP00}.

In the present paper we propose an extension of the scalar mode solver
\cite{IHS07} to vectorial problems. Our method is based on
individual expansions of each mode profile component into a set of
a priori defined functions of one coordinate axis (vertical),
here, field components of some slab waveguide mode. The expansion
is global, meaning that the same basis functions are used at any
point on the horizontal axis. The unknown expansion coefficients
-- in our case functions, defined on the horizontal axis -- are
found by means of variational methods \cite{Vas91,GrM07}.

The present method can be viewed as some bridge between two
popular approaches, namely the FMM on the one hand and the EIM on
the other. In the standard EIM the 2D problem of finding modes of
the waveguide is reduced to consecutive solving two 1D problems:
at first, the 1D modes, and their propagation constants, of the
constituting slab waveguides are found, and then their propagation
constants are used to define effective refractive indices of a
reduced 1D problem. In general this is a very quick and easy
approach for a rough estimation of mode parameters. However, in
case one of the constituting slabs doesn't support a guided mode
(for example, some substrate material with air on top) it is
impossible to uniquely define the effective refractive index in
that particular region of the reduced problem. Should it be the
refractive index of the cladding, refractive index of the air, or
something in between? The restriction of the present approach to
one-term expansions will answer this question.

The validity of the method was checked on several structures,
including waveguides with rectangular and non-rectangular
piecewise-constant refractive index distributions, and a diffused
waveguide. Comparison shows that the present method is a more
consistent and accurate alternative to the standard EIM and also
can be pushed to its limits and used for rigorous computations.

The paper is organized as follows. In \autoref{sec:Variational
Form of the Vectorial Mode Problem} the problem of finding
vectorial modes of the dielectric waveguides is stated, then some
properties of slab modes and the modal field ansatz are described
in sections \ref{sec:Slab Modes} and \ref{sec:Modal Field Ansatz}.
The equations for the coefficient functions are derived in \autoref{sec:Reduced
problem}. Section \ref{sec:Method of solution} outlines the
numerical solution methods. The relation of the present method to
the EIM and FMM is explained in more detail in sections
\ref{sec:Connection to the Effective Index Method} and
\ref{sec:Connection to the Film Mode Matching Method}. Then in
\autoref{sec:Numerical Results} numerical results for several
waveguide configurations are presented. Finally some concluding
remarks are made in \autoref{sec:Concluding Remarks}.

\section{Variational form of the vectorial mode problem}
\label{sec:Variational Form of the Vectorial Mode Problem}

Consider a $z$-invariant dielectric isotropic waveguide defined on
its cross-section by a refractive index $n(x,y)$ or
relative dielectric permittivity $\varepsilon(x,y) = n^2(x,y)$ distribution.
\reffig{Introduction} shows two examples.

\FigSideHeight{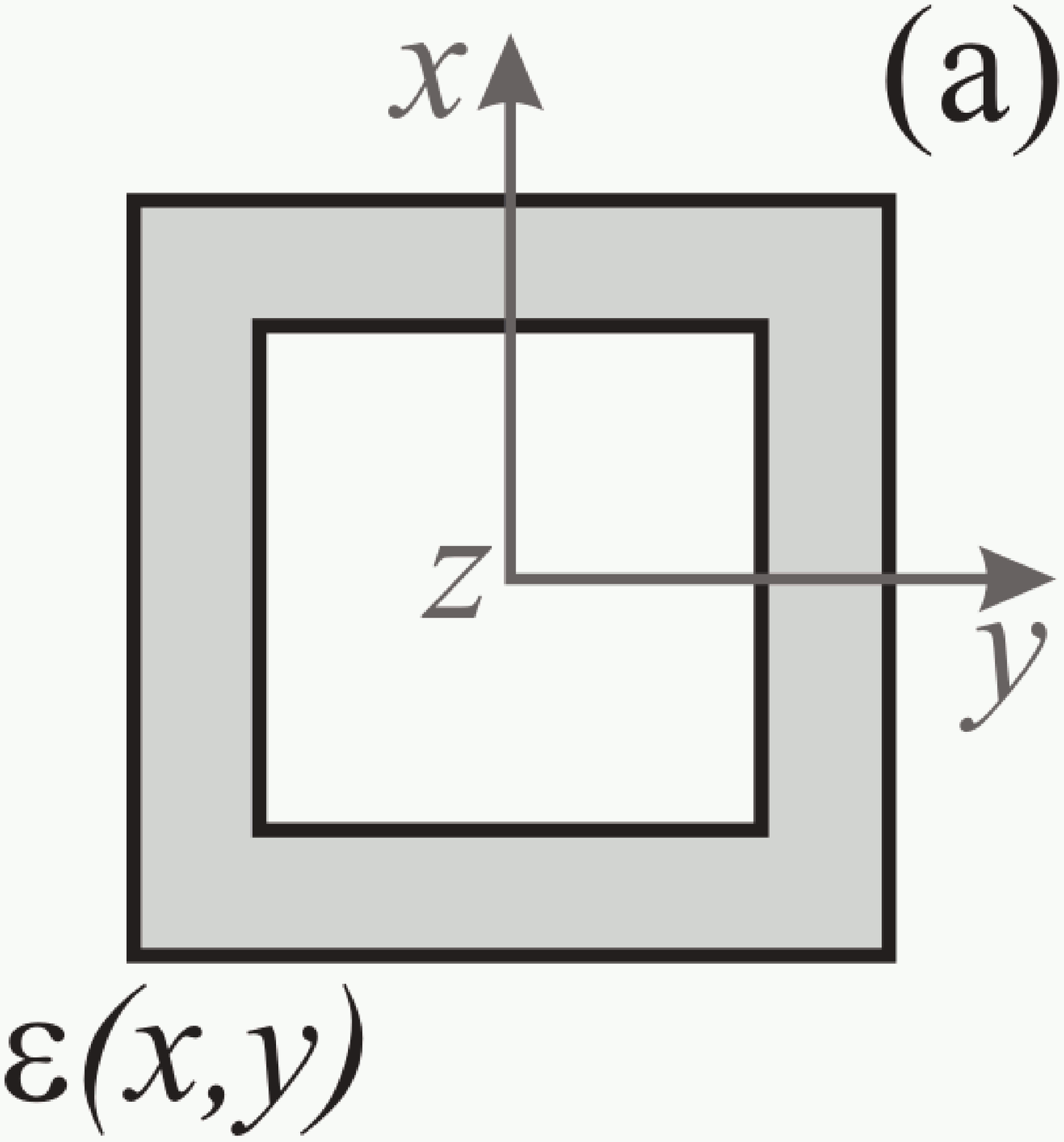}{2.5cm}{Examples for 3D dielectric
waveguides defined on their cross-section by permittivity
distribution $\varepsilon(x,y)$. The structures are invariant
along the $z$-axis. (a) box-shaped hollow-core waveguide, a
concept from \cite{MMM07}, the subject of section
\ref{sec:Waveguide with piecewise rectangular cross-section}, (b)
a standard rib waveguide, investigated in section
\ref{sec:Rectangular Rib Waveguide}.}
\noindent The propagation of monochromatic light, given by the
electric $\bar{\mathbf{E}}$ and magnetic $\bar{\mathbf{H}}$
components of the optical field, with propagation constant $\beta$
and frequency $\omega$,
\begin{equation}
\bar{\mathbf{E}}(x,y,z,t) = \mathbf{E}(x,y) \nexp{-\iu \beta z}
\nexp{\iu \omega t}, \quad \bar{\mathbf{H}}(x,y,z,t) =
\mathbf{H}(x,y) \nexp{-\iu \beta z} \nexp{\iu \omega t},
\end{equation}
is governed by the Maxwell equations for the mode profile
components $\mathbf{E}$ and $\mathbf{H}$
\begin{equation}
\begin{array}{l}
\omega \varepsilon_0 \varepsilon \mathbf{E} + \iu C \mathbf{H} = \beta R \mathbf{H}, \\
\omega \mu_0 \mu \mathbf{H} - \iu C \mathbf{E} = - \beta R
\mathbf{E},
\end{array}
\label{Maxwell Equations}
\end{equation}
with
\begin{equation}
R = \left(%
\begin{array}{ccc}
  0 & 1 & 0 \\
  -1 & 0 & 0 \\
  0 & 0 & 0 \\
\end{array}%
\right), \qquad C = \left(%
\begin{array}{ccc}
  0 & 0 & \partial_y \\
  0 & 0 & -\partial_x \\
  -\partial_y & \partial_x & 0 \\
\end{array}%
\right), \label{R and C matrices}
\end{equation}
vacuum permittivity $\varepsilon_0$, vacuum permeability $\mu_0$,
relative permittivity $\varepsilon(x,y) = n^2(x,y)$.
Here and further in this paper it is assumed that the relative
permeability $\mu$ is equal to $1$, as is the case for most
materials at optical frequencies.

We will work with a variational formulation of the Maxwell
equations. Solutions $(\beta, \mathbf{E}, \mathbf{H})$ of the
equations (\ref{Maxwell Equations}) correspond to stationary
points $(\mathbf{E}, \mathbf{H})$ of the functional
\cite{Vas91}
\begin{equation}
\mathcal{F}(\mathbf{E}, \mathbf{H})= \frac{\omega \varepsilon_0
\langle \mathbf{E}, \varepsilon \mathbf{E} \rangle + \omega \mu_0
\langle \mathbf{H}, \mathbf{H} \rangle + \iu \langle \mathbf{E}, C
\mathbf{H} \rangle - \iu \langle \mathbf{H}, C \mathbf{E}
\rangle}{ \langle \mathbf{E}, R \mathbf{H} \rangle - \langle
\mathbf{H}, R \mathbf{E} \rangle}, \label{vms:Functional}
\end{equation}
with propagation constant $\beta = \mathcal{F}(\mathbf{E},
\mathbf{H})$ equal to the value of the functional at the
stationary point. The inner product used is $\langle \mathbf{A},
\mathbf{B}\rangle = \int \mathbf{A}^* \cdot \mathbf{B} \,dx \,dy$.
The natural interface conditions are the continuity of all
tangential field components across the interfaces.

\section{Slab modes}
\label{sec:Slab Modes} In this section we will consider modes of
slab waveguides, which we will use in the next section as building blocks to
construct approximations of the modes of waveguides with
arbitrary 2D cross-sections.
Furthermore, we introduce rotations of the slab modes; these rotations will be
needed to provide a physical motivation for the particular form of the approximations
that we will employ.

\FigSideHeight{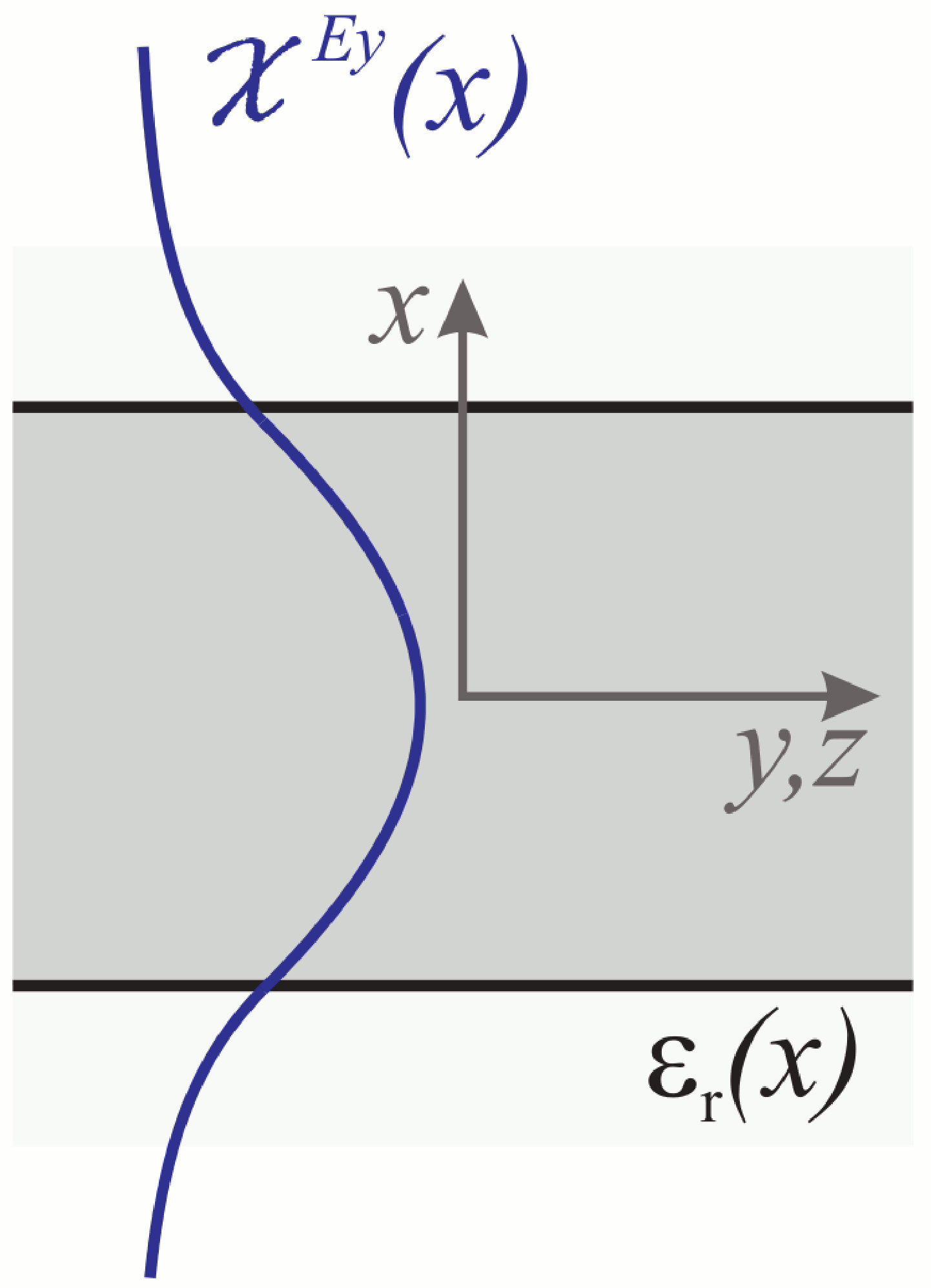}{4.0cm}{A slab waveguide with
permittivity distribution $\varepsilon_r(x)$ and principal
component $\chi^{E_y}$ of a corresponding TE slab mode
(\ref{vms:TE slab mode}).}

A one dimensional TE mode, propagating in the $z$-direction with
propagation constant $\beta_{{\text{r},}_{\text{TE}}}$, of the
slab waveguide, given by the permittivity distribution
$\varepsilon_{\text{r}}(x)$ (\reffig{SlabModes}) can be
represented as
\begin{equation}
\begin{array}{c}
\left(
\begin{array}{c}
  E_x, E_y, E_z \\
  H_x, H_y, H_z
\end{array}
\right)(x,y,z) = \left(
\begin{array}{ccc}
   0, & \chi^{E_y}(x) , & 0 \\
   \chi^{H_x}(x), & 0, & \chi^{H_z}(x)
\end{array}
\right) \nexp{ - \iu \beta_{{\text{r},}_{\text{TE}}}z}.
\end{array}
\label{vms:TE slab mode}
\end{equation}
The principal electric component $\chi^{E_y}$ satisfies the
equation
\begin{equation}
\left(\chi^{E_y}(x) \right)'' + k^2 \varepsilon_{\text{r}}(x)
\chi^{E_y}(x) = \beta_{{\text{r},}_{\text{TE}}}^2 \chi^{E_y}(x)
\end{equation}
\noindent with vacuum wavenumber $k = 2\pi/\lambda$. The remaining
two nonzero components of the mode profile can be derived directly
from $\chi^{E_y}$:

\begin{equation}
\chi^{H_x}(x) = - \frac{\beta_{{\text{r},}_{\text{TE}}}}{\omega \mu_0} \chi^{E_y}(x), \qquad
\chi^{H_z}(x) = \frac{i}{\omega \mu_0} \left( \chi^{E_y}(x) \right)'.
\label{XHx and XHz through XEy}
\end{equation}
The slab waveguide (\reffig{SlabModes}) is by definition invariant
in the $(y,z)$-plane. So if a modal solution of Maxwell equations
propagating in the $z$-direction will be rotated in the
$(y,z)$-plane by an angle $\theta$
(\reffig{RotationAngleFieldTETM}), it will still remain a modal
solution of the Maxwell equations, but now propagating in the
direction $(y, z) = (- \sin \theta; \cos \theta)$:
\begin{equation}
\begin{array}{l}
\left(
\begin{array}{c}
  E_x, E_y, E_z \\
  H_x, H_y, H_z
\end{array}
\right)(x,y,z) =
\left(
\begin{array}{ccc}
   0, & \chi^{E_y}(x) \cos \theta, & \chi^{E_y}(x) \sin \theta \\
   \chi^{H_x}(x), & - \chi^{H_z}(x) \sin \theta, & \chi^{H_z}(x) \cos \theta
\end{array}
\right) \nexp{ - \iu \beta_{{\text{r},}_{\text{TE}}} ( - \sin
\theta y + \cos \theta z)}.
\end{array}
\label{vms:rotated TE slab mode}
\end{equation}

\FigSideHeight{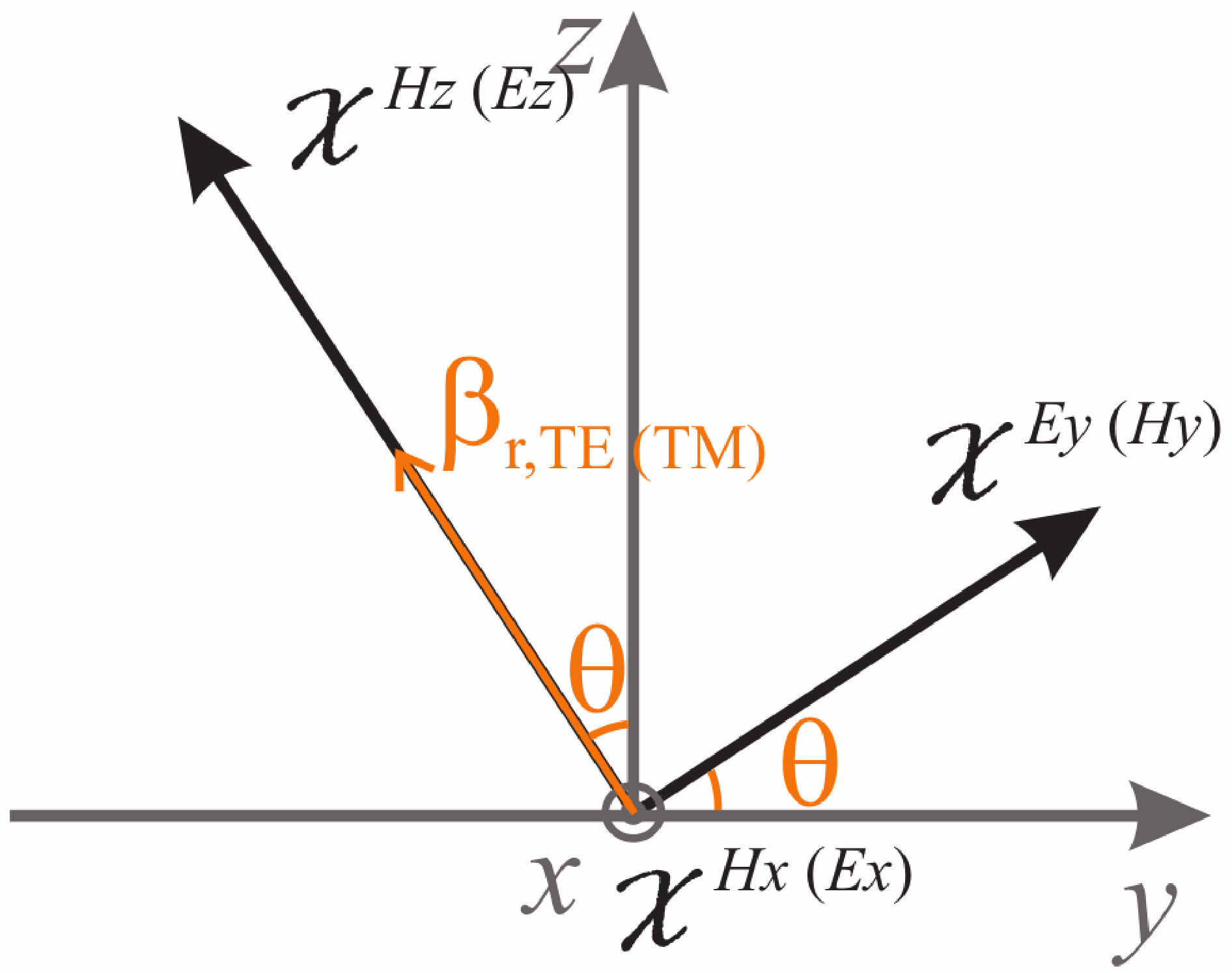}{3.0cm}{A slab TE (TM) mode
propagating in the $z$-direction with propagation constant
$\beta_{\text{r,TE (TM)}}$ is rotated around the $x$-axis by an
angle $\theta$. The rotated mode propagates with the same
propagation constant, but in the direction $(y,z)=(-\sin \theta,
\cos \theta)$.}

Similarly a 1D TM slab mode, propagating in the $z$-direction with
propagation constant $\beta_{{\text{r},}_{\text{TM}}}$
\begin{equation}
\begin{array}{c}
\left(
\begin{array}{c}
  E_x, E_y, E_z \\
  H_x, H_y, H_z
\end{array}
\right)(x,y,z) = \left(
\begin{array}{ccc}
   \chi^{E_x}(x), & 0 , & \chi^{E_z}(x) \\
   0, & \chi^{H_y}(x), & 0
\end{array}
\right) \nexp{ - \iu \beta_{{\text{r},}_{\text{TM}}} z},
\end{array}
\label{vms:TM slab mode}
\end{equation}
will still be a solution of the Maxwell equations after a rotation
around the $x$-axis (\reffig{RotationAngleFieldTETM})
\begin{equation}
\begin{array}{l}
\left(
\begin{array}{c}
  E_x, E_y, E_z \\
  H_x, H_y, H_z
\end{array}
\right)(x,y,z) =
\left(
\begin{array}{ccc}
   \chi^{E_x}(x), & - \chi^{E_z}(x) \sin \theta, & \chi^{E_z}(x) \cos \theta \\
   0, & \chi^{H_y}(x) \cos \theta, & \chi^{H_y}(x) \sin \theta
\end{array}
\right) \nexp{ - \iu \beta_{{\text{r},}_{\text{TM}}} ( - \sin
\theta y + \cos \theta z)}.
\end{array}
\label{vms:rotated TM slab mode}
\end{equation}
The principal magnetic component $\chi^{H_y}$ satisfies the
equation
\begin{equation}
\left(\frac{1}{\varepsilon_{\text{r}}(x)} (\chi^{H_y}(x))'
\right)' + k^2 \chi^{H_y}(x) = \beta_{{\text{r},}_{\text{TM}}}^2
\frac{1}{\varepsilon_{\text{r}}(x)} \chi^{H_y}(x).
\end{equation}
Again the remaining two nonzero components of the mode profile can
be derived directly from $\chi^{H_y}$:
\begin{equation}
\chi^{E_x}(x) = \frac{\beta_{{\text{r},}_{\text{TM}}}}{\omega \varepsilon_0
\varepsilon_{\text{r}}(x)} \chi^{H_y}(x), \qquad \chi^{E_z}(x) = -
\frac{i}{\omega \varepsilon_0 \varepsilon_{\text{r}}(x)} \left(
\chi^{H_y}(x) \right)'. \label{XEx and XEz through XHy}
\end{equation}

\section{Modal field ansatz}
\label{sec:Modal Field Ansatz}

We now return to the vectorial modes of the 3D waveguides, as in
section \ref{sec:Variational Form of the Vectorial Mode Problem}.
Each field component $F \in \{E_x, E_y, E_z, H_x, H_y, H_z \}$ is
represented individually as a superposition of $m_F$ a priori
known functions $X^{F}_j(x)$, defined on one coordinate axis,
times some unknown coefficient-function $Y^{F}_j (y)$, defined on
the other axis:
\begin{equation}
F(x,y) = \sum \limits_{j=1}^{m_F} X^{F}_j (x) Y^{F}_j (y).
\label{FieldTemplate}
\end{equation}

For the functions $X$ we will take components of slab modes from
some reference slice(s). Further in the paper two types of the
expansion will be relevant, one which introduces 5 unknown
functions $Y$ per slab mode, and another one, which introduces
only 3. These will be
called five component approximation (VEIM5) and three component
approximation (VEIM3), respectively.

In case of VEIM5, the TE basis mode (\ref{vms:TE
slab mode}) number\ $j$ with mode profile components $\chi^{E_y}_j$,
$\chi^{H_x}_j$, $\chi^{H_z}_j$ contributes to the expansion of
components $E_y$, $E_z$, $H_x$, $H_y$ and $H_z$ with the form
$\chi^{E_y}_j Y^{E_y}_j$, $\chi^{E_y}_j Y^{E_z}_j$, $\chi^{H_x}_j
Y^{H_x}_j$, $\chi^{H_z}_j Y^{H_y}_j$ and $\chi^{H_z}_j Y^{H_z}_j$.
Likewise, the TM basis mode (\ref{vms:TM slab mode}) number\ $l$ with
mode profile components $\chi^{E_y}_l$, $\chi^{H_x}_l$,
$\chi^{H_z}_l$ contributes to the expansion of components $E_x$,
$E_y$, $E_z$, $H_y$ and $H_z$ with the form $\chi^{E_x}_l
Y^{E_x}_l$, $\chi^{E_z}_l Y^{E_y}_l$, $\chi^{E_z}_l Y^{E_z}_l$,
$\chi^{H_y}_l Y^{H_y}_l$, $\chi^{H_y}_l Y^{H_z}_l$, such that the
complete expansion looks like
\begin{equation}
\begin{array}{r}
\left(
\begin{array}{c}
  E_x, E_y, E_z \\
  H_x, H_y, H_z
\end{array}
\right)(x,y,z)
=
\sum \limits_{j \in \text{TE}} \left(
\begin{array}{ccc}
   0, & \chi^{E_y}_j(x) Y^{E_y}_j(y), & \chi^{E_y}_j(x) Y^{E_z}_j(y) \\
   \chi^{H_x}_j(x) Y^{H_x}_j(y), & \chi^{H_z}_j(x) Y^{H_y}_j(y), & \chi^{H_z}_j(x)
   Y^{H_z}_j(y)
\end{array}
\right) + \vspace{0.2cm}\\
+ \sum \limits_{l \in \text{TM}} \left(
\begin{array}{ccc}
   \chi^{E_x}_l(x) Y^{E_x}_l(y), & \chi^{E_z}_l(x) Y^{E_y}_l(y), & \chi^{E_z}_l(x) Y^{E_z}_l(y) \\
   0, & \chi^{H_y}_l(x) Y^{H_y}_l(y), & \chi^{H_y}_l(x) Y^{H_z}_l(y)
\end{array}
\right).
\end{array}
\label{5CA}
\end{equation}

This expansion has the drawback that the functions making up some
of the components can become linearly dependent; for example, the full set of $\chi^{H_y}_l(x)$ components from TM modes form a complete set; thus any $\chi^{H_z}_j(x)$ from a TE mode can be expressed in that complete set of functions. When using a limited number of modes in the expansion, no problems result from this; however, increasing the number of modes will at some point make the problem ill-conditioned. Therefore, we introduce a different expansion, which we call VEIM3, in which we omit contributions of some modal components - making sure that each vector component is only represented by either TE of TM slab mode components. So a TE basis mode (\ref{vms:TE slab mode}) number\ $j$
with mode profile components $\chi^{E_y}_j$, $\chi^{H_x}_j$,
$\chi^{H_z}_j$ contributes to the expansion of components $E_y$,
$E_z$ and $H_x$ with the form $\chi^{E_y}_j Y^{E_y}_j$,
$\chi^{E_y}_j Y^{E_z}_j$, $\chi^{H_x}_j Y^{H_x}_j$. Likewise a TM
basis mode (\ref{vms:TM slab mode}) number\ $l$ with mode profile
components $\chi^{E_y}_l$, $\chi^{H_x}_l$, $\chi^{H_z}_l$
contributes to the expansion of components $E_x$, $H_y$ and $H_z$
with the form $\chi^{E_x}_l Y^{E_x}_l$, $\chi^{H_y}_l Y^{H_y}_l$,
$\chi^{H_y}_l Y^{H_z}_l$, such that the complete expansion looks
like
\begin{equation}
\begin{array}{r}
\left(
\begin{array}{c}
  E_x, E_y, E_z \\
  H_x, H_y, H_z
\end{array}
\right)(x,y,z)
=
\sum \limits_{j \in \text{TE}} \left(
\begin{array}{ccc}
   0, & \chi^{E_y}_j(x) Y^{E_y}_j(y), & \chi^{E_y}_j(x) Y^{E_z}_j(y) \\
   \chi^{H_x}_j(x) Y^{H_x}_j(y), & 0, & 0
\end{array}
\right) + \vspace{0.2cm} \\
+ \sum \limits_{l \in \text{TM}} \left(
\begin{array}{ccc}
   \chi^{E_x}_l(x) Y^{E_x}_l(y), & 0, & 0 \\
   0, & \chi^{H_y}_l(x) Y^{H_y}_l(y), & \chi^{H_y}_l(x) Y^{H_z}_l(y)
\end{array}
\right).
\end{array}
\label{3CA}
\end{equation}

Note that in both expansions each contributing component $\chi^F$
of a 1D mode is used to represent the field not only in the slab
segment where it belongs, as in EIM and FMM methods, but also
in the whole waveguide. So even with a single slab mode in both
expansions, (\ref{5CA}) and (\ref{3CA}), it is possible to
construct an approximation of the field in the whole structure. In
section \ref{sec:Connection to the Effective Index Method} we will
study in detail properties of such one-mode-expansions.

The form of the expansion (\ref{5CA}) was inspired by the mode
matching techniques that use the physically motivated approach of employing rotated modes (\ref{vms:rotated TE
slab mode}), (\ref{vms:rotated TM slab mode}) to locally expand
the total field \cite{Bie04a,Sud93b}. In the present approach
though, we attribute those parts of the slab mode components that
do not depend on $x$ to the functions $Y^F$, treating them as
unknowns -- but the $x$-dependence of the $y$ and $z$ components is still the same. In the sections \ref{sec:Connection to the Effective
Index Method} and \ref{sec:Connection to the Film Mode Matching
Method} we will study the behavior of these functions $Y^F$.

What concerns the choice of the reference slice(s), it seems that
modal components from the slice, where the maximum power is
expected to be localized, give the best results. Further in this
paper VEIM5 will be used with a few modes only
for rough and efficient approximations, while
VEIM3 will be used with higher numbers of modes
to obtain accurate, converged results.
We do not restrict to using modes from one reference slice only;
we observe that adding mode(s) from another slice can greatly
improve accuracy for lower number of modes in the expansion. However,
care must be taken; the problem can become ill-conditioned if the
modes become nearly linear dependent.

In the following all the slab mode components $\chi$, which are
used to expand a field component $F$ of the complete waveguide, we
will denote as $X^F$ (just like in eqn.\ (\ref{FieldTemplate})).

\section{Reduced problem}
\label{sec:Reduced problem}

The next question is how to find corresponding functions $Y$, such
that the expansion (\ref{FieldTemplate}) represents the true
solution in the best possible way. For this purpose we apply
variational restriction \cite{Vas97,GrM07} of
the functional (\ref{vms:Functional}). In short it can be outlined
as follows. As it was already mentioned the critical points of the
functional (\ref{vms:Functional}), which satisfy some continuity
conditions, are solutions of the Maxwell equations (\ref{Maxwell
Equations}) and, vice versa, solutions of the Maxwell equations
(\ref{Maxwell Equations}) are critical points of the functional
(\ref{vms:Functional}).

After insertion of the expansions (\ref{5CA}) or (\ref{3CA}),
variation of the functional (\ref{vms:Functional}) with respect to
a function $\mathbf{Y}^F$, a vector function made up of all
functions $Y^F$, results in the following system of first order
differential equations for $\mathbf{Y}^F$ with parameter $\beta$:
\begin{equation}
\begin{array}{l}
 \mathbf{A}_{11} \mathbf{Y}^{E_x} + \mathbf{A}_{12} (\mathbf{Y}^{H_z})' = \beta \mathbf{A}_{13} \mathbf{Y}^{H_y} \\

 \mathbf{A}_{21} \mathbf{Y}^{E_y} + \mathbf{A}_{22} \mathbf{Y}^{H_z} = \beta \mathbf{A}_{23} \mathbf{Y}^{H_x} \\

 \mathbf{A}_{31} \mathbf{Y}^{E_z} + \mathbf{A}_{32} (\mathbf{Y}^{H_x})' + \mathbf{A}_{33} \mathbf{Y}^{H_y} = 0 \\

 \mathbf{A}_{41} \mathbf{Y}^{H_x} + \mathbf{A}_{42} (\mathbf{Y}^{E_z})' = \beta \mathbf{A}_{43} \mathbf{Y}^{E_y} \\

 \mathbf{A}_{51} \mathbf{Y}^{H_y} + \mathbf{A}_{52} \mathbf{Y}^{E_z} = \beta \mathbf{A}_{53} \mathbf{Y}^{E_x}\\

 \mathbf{A}_{61} \mathbf{Y}^{H_z} + \mathbf{A}_{62} (\mathbf{Y}^{E_x})' + \mathbf{A}_{63} \mathbf{Y}^{E_y} = 0.
\end{array}
\label{EquationsForYFunctions}
\end{equation}
\noindent The elements of the matrices $\mathbf{A}$ include the
overlap integrals (here: $\langle a, b \rangle = \int a^* b \,dx$) of the functions $X^F_j(x)$, their derivatives,
and the local permittivity distribution of the waveguide:
\begin{equation}
\begin{array}{lll}
 \mathbf{A}_{11} (p,j) = \omega \langle X_p^{E_x}, \varepsilon X_j^{E_x} \rangle &
 \mathbf{A}_{12}(p,j) = \iu \langle X_p^{E_x}, X_j^{H_z} \rangle &
 \mathbf{A}_{13}(p,j) = \langle X_p^{E_x}, X_j^{H_y} \rangle \\

 \mathbf{A}_{21}(p,j) = \omega \langle X_p^{E_y}, \varepsilon X_j^{E_y} \rangle &
 \mathbf{A}_{22}(p,j) = - \iu \langle X_p^{E_y}, (X_j^{H_z})' \rangle &
 \mathbf{A}_{23}(p,j) = - \langle X_p^{E_y}, X_j^{H_x} \rangle \\

 \mathbf{A}_{31}(p,j) = \omega \langle X_p^{E_z}, \varepsilon X_j^{E_z} \rangle &
 \mathbf{A}_{32}(p,j) = - \iu \langle X_p^{E_z}, X_j^{H_x} \rangle &
 \mathbf{A}_{33}(p,j) = \iu \langle X_p^{E_z}, (X_j^{H_y})' \rangle \\

 \mathbf{A}_{41}(p,j) = \omega \mu \langle X_p^{H_x}, X_j^{H_x} \rangle &
 \mathbf{A}_{42}(p,j) = - \iu \langle X_p^{H_x}, X_j^{E_z} \rangle &
 \mathbf{A}_{43}(p,j) = - \langle X_p^{H_x}, X_j^{E_y} \rangle \\

 \mathbf{A}_{51}(p,j) = \omega \mu \langle X_p^{H_y}, X_j^{H_y} \rangle &
 \mathbf{A}_{52}(p,j) = \iu \langle X_p^{H_y}, (X_j^{E_z})' \rangle &
 \mathbf{A}_{53}(p,j) = \langle X_p^{H_y}, X_j^{E_x} \rangle \\

 \mathbf{A}_{61}(p,j) = \omega \mu \langle X_p^{H_z}, X_j^{H_z} \rangle &
 \mathbf{A}_{62}(p,j) = \iu \langle X_p^{H_z}, X_j^{E_x} \rangle &
 \mathbf{A}_{63}(p,j) = - \iu \langle X_p^{H_z}, (X_j^{E_y})' \rangle
\end{array}
\label{Matrices A}
\end{equation}
%
\noindent Note that the permittivity appears only in
$\mathbf{A}_{11}$, $\mathbf{A}_{21}$ and $\mathbf{A}_{31}$, hence
only these matrices are $y$-dependent.

If the permittivity exhibits discontinuities along the
$y$-direction, the functions
\begin{equation}
\mathbf{Y}^{E_x} \quad \text{and} \quad \mathbf{Y}^{H_x}, \qquad
\mathbf{Y}^{E_z} \quad \text{and} \quad \mathbf{Y}^{H_z}
\label{vms: component_continuity_conditions}
\end{equation}
are required to be continuous at the respective positions.

It turns out that by algebraic operations the
system of first order differential equations
(\ref{EquationsForYFunctions}) can be reduced to a system of
second order differential equations for the vector functions
$\mathbf{Y}^{E_x}$ and $\mathbf{Y}^{H_x}$ only. Moreover, since
the components $E_y$ and $E_z$, $H_y$ and $H_z$ are approximated by
the same functions $\chi$ in the representations (\ref{5CA}) and
(\ref{3CA}), the matrices $\mathbf{A}$ satisfy
the following equalities:
\begin{equation}
\begin{array}{l}
 \mathbf{A}_{13} = - \iu \mathbf{A}_{12} \\
 \mathbf{A}_{31} = \mathbf{A}_{21}, \quad \mathbf{A}_{32} = \iu \mathbf{A}_{23}, \quad \mathbf{A}_{33} = - \mathbf{A}_{22} \\
 \mathbf{A}_{43} = - \iu \mathbf{A}_{42} \\
 \mathbf{A}_{61} = \mathbf{A}_{51}, \quad \mathbf{A}_{62} = \iu \mathbf{A}_{53}, \quad \mathbf{A}_{63} = - \mathbf{A}_{52},
\end{array}
\end{equation}
and hence the system (\ref{EquationsForYFunctions}) reduces to
\begin{equation}
\mathbf{S}_1 \mathbf{u} + (\mathbf{S}_2 \mathbf{u}' + \beta
\mathbf{S}_3 \mathbf{u})' = \beta^2 \mathbf{S}_2 \mathbf{u} +
\beta \mathbf{S}_3 \mathbf{u}', \label{vms:System For YEx and YHx}
\end{equation}
with $\mathbf{u}(y) = \left(%
\begin{array}{c}
  \mathbf{Y}^{E_x}(y) \\
  \mathbf{Y}^{H_x}(y) \\
\end{array}%
\right)$ and (anti-)block-diagonal matrices $\mathbf{S}$ of the
following form:
\begin{equation}
\begin{array}{l}

 \mathbf{S}_1 = \left(%
\begin{array}{cc}
  \mathbf{A}_{11} & 0 \\
  0 & \mathbf{A}_{41} \\
\end{array}%
\right), \vspace{0.2cm} \\
\mathbf{S}_2 =
 \left( \begin{array}{cc}
  - \iu \mathbf{A}_{12} \big( \mathbf{A}_{51} + \mathbf{A}_{52} \mathbf{A}_{21}^{-1} \mathbf{A}_{22} \big)^{-1} \mathbf{A}_{53} & 0 \\
  0 & - \iu \mathbf{A}_{42} \big( \mathbf{A}_{21} + \mathbf{A}_{22} \mathbf{A}_{51}^{-1} \mathbf{A}_{52} \big)^{-1} \mathbf{A}_{23} \\
\end{array}%
\right), \vspace{0.2cm} \\
\mathbf{S}_3 =
 \left(\begin{array}{cc}
  0 & \mathbf{A}_{12} \mathbf{A}_{51}^{-1} \mathbf{A}_{52} \big( \mathbf{A}_{21} + \mathbf{A}_{22} \mathbf{A}_{51}^{-1} \mathbf{A}_{52} \big)^{-1} \mathbf{A}_{23} \\
  \mathbf{A}_{42} \mathbf{A}_{21}^{-1} \mathbf{A}_{22} \big( \mathbf{A}_{51} + \mathbf{A}_{52} \mathbf{A}_{21}^{-1} \mathbf{A}_{22} \big)^{-1} \mathbf{A}_{53} & 0 \\
\end{array}%
\right).
\end{array}
\label{Matrices S simplified}
\end{equation}
%
\noindent Across the vertical interfaces continuity of
\begin{equation}
\mathbf{u} \quad \text{and} \quad \mathbf{S}_2 \mathbf{u}' +\beta
\mathbf{S}_3 \mathbf{u}
\label{ContinuityConditionsAcrossTheInterfaces}
\end{equation}
is required.

As soon as the function $\mathbf{u}$, or in other words
$\mathbf{Y}^{E_x}$ and $\mathbf{Y}^{H_x}$, are known, the
functions $\mathbf{Y}$ corresponding to the four other components
can be derived as follows:
\begin{equation}
\begin{array}{l}
\mathbf{Y}^{E_y} = \iu \mathbf{A}_{21}^{-1} \mathbf{A}_{22} \big( \mathbf{A}_{51} + \mathbf{A}_{52} \mathbf{A}_{21}^{-1} \mathbf{A}_{22} \big)^{-1} \mathbf{A}_{53} (\mathbf{Y}^{E_x})' + \beta \big( \mathbf{A}_{21} + \mathbf{A}_{22} \mathbf{A}_{51}^{-1} \mathbf{A}_{52} \big)^{-1} \mathbf{A}_{23} \mathbf{Y}^{H_x}, \\
\mathbf{Y}^{E_z} = \beta \mathbf{A}_{21}^{-1} \mathbf{A}_{22} \big( \mathbf{A}_{51} + \mathbf{A}_{52} \mathbf{A}_{21}^{-1} \mathbf{A}_{22} \big)^{-1} \mathbf{A}_{53} \mathbf{Y}^{E_x} - \iu \big( \mathbf{A}_{21} + \mathbf{A}_{22} \mathbf{A}_{51}^{-1} \mathbf{A}_{52} \big)^{-1} \mathbf{A}_{23} (\mathbf{Y}^{H_x})', \\
\mathbf{Y}^{H_y} = \beta \big( \mathbf{A}_{51} + \mathbf{A}_{52} \mathbf{A}_{21}^{-1} \mathbf{A}_{22} \big)^{-1} \mathbf{A}_{53} \mathbf{Y}^{E_x} + \iu \mathbf{A}_{51}^{-1} \mathbf{A}_{52} \big( \mathbf{A}_{21} + \mathbf{A}_{22} \mathbf{A}_{51}^{-1} \mathbf{A}_{52} \big)^{-1} \mathbf{A}_{23} (\mathbf{Y}^{H_x})', \\
\mathbf{Y}^{H_z} = - \iu \big( \mathbf{A}_{51} + \mathbf{A}_{52}
\mathbf{A}_{21}^{-1} \mathbf{A}_{22} \big)^{-1} \mathbf{A}_{53}
(\mathbf{Y}^{E_x})' + \beta \mathbf{A}_{51}^{-1} \mathbf{A}_{52}
\big( \mathbf{A}_{21} + \mathbf{A}_{22} \mathbf{A}_{51}^{-1}
\mathbf{A}_{52} \big)^{-1} \mathbf{A}_{23} \mathbf{Y}^{H_x}.
\end{array}
\label{vms:Components Ey, Ez, Hy and Hz}
\end{equation}

Note that substituting Equation (\ref{vms:Components Ey, Ez, Hy and Hz}) into the continuity conditions (\ref{ContinuityConditionsAcrossTheInterfaces}) shows that (\ref{ContinuityConditionsAcrossTheInterfaces}) exactly implies the continuity of the relevant electromagnetic components (\ref{vms: component_continuity_conditions}).

\section{Method of solution}
\label{sec:Method of solution}

In general the system (\ref{vms:System For YEx and YHx}) can be
solved by the Finite Element method \cite{COMSOL, KSV05}. It relies on a spatial
discretization, i.e.\ divides the whole computational domain into
a number of elements. On each of these elements the unknown
function is represented as a superposition of some basis
functions. The coefficients of the expansion are found using the
weak form of eqn.\ (\ref{vms:System For YEx and YHx}). While this
method is very general, it quickly introduces a large number of
unknowns.

However, due to common techniques of fabrication many waveguides
do not have a completely arbitrary refractive index distribution,
but rather one which is piecewise constant along the horizontal
axis. The waveguide then can be split in several vertical slices,
where the refractive index does not change in the horizontal
direction. In each of these layers the general solution of
(\ref{vms:System For YEx and YHx}) can be written down
analytically. Gluing them together across the vertical interfaces
will give the desired mode profile.

Both of these methods can be applied to find not only the fundamental,
but also higher order modes. In the following we will outline each of these methods in more
detail.

\subsection{Arbitrary refractive index distribution: Finite Element Method}
\label{sec:Waveguides with arbitrary refractive index
distribution}

In case of an arbitrary permittivity distribution
$\varepsilon(x,y)$ (diffused waveguide, waveguide with slanted
sidewalls) the matrices $\mathbf{S}$ depend on $y$, as their
elements include overlap integrals with the permittivity
$\varepsilon(x,y)$. One of the ways to solve the differential
equation (\ref{vms:System For YEx and YHx}) is by using the Finite
Element Method.

By multiplying both sides of (\ref{vms:System For YEx and YHx})
from the left by some continuous test vector-function $\mathbf{v}$
and integrating over $y$ one gets the weak form of equation
(\ref{vms:System For YEx and YHx}):
\begin{equation}
 \int \big( - \mathbf{v}^\intercal \mathbf{S}_1 \mathbf{u} + (\mathbf{v}^\intercal)'
\mathbf{S}_2 \mathbf{u}' \big) \,dy + \beta \int \big(
(\mathbf{v}^\intercal)' \mathbf{S}_3 \mathbf{u} +
\mathbf{v}^\intercal \mathbf{S}_3 \mathbf{u}' \big) \,dy + \beta^2
\int \mathbf{v}^\intercal \mathbf{S}_2 \mathbf{u} \,dy = 0.
\label{WeakFormulation}
\end{equation}
Then we expand the solution $\mathbf{u}$ into a finite combination
of the basis functions $\boldsymbol{\varphi}_{ij}$,
\begin{equation}
\mathbf{u}(y) = \sum \limits_{i=1}^{n_d} \sum \limits_{j=1}^{n_g}
a_{ij} \boldsymbol{\varphi}_{ij}(y), \label{expansion}
\end{equation}
\noindent with $n_d$ the dimension of the vector $\mathbf{u}$;
$n_g$ the number of consecutive grid points $y_j$ into which the
$y$-axis has been divided, and
\begin{equation}
\boldsymbol{\varphi}_{ij}(y) = \left(%
\begin{array}{c}
  0 \\
  \vdots \\
  \hat{\varphi}_j(y) \\
  \vdots \\
  0 \\
\end{array}%
\right) \ \leftarrow \ i^{\text{th}} \ \text{position}
\label{Vectorial Basis Function}
\end{equation}
with, for example, linear basis functions 
\begin{equation}
\hat{\varphi}_j(y) = \left\{%
\begin{array}{ll}
    0, & y < y_{j-1} \ \text{or} \ y \geq y_{j+1}; \vspace{0.2cm}\\
    \frac{y - y_{j-1}}{y_j - y_{j-i}}, & y_{j-1} \leq y < y_j; \vspace{0.2cm} \\
    \frac{y_{j+1} - y}{y_{j+1} - y_j}, & y_{j} \leq y < y_{j+1}. \\
\end{array}%
\right. \label{Basis Function per Element}
\end{equation}

As eqn.\ (\ref{WeakFormulation}) should hold for an arbitrary
continuous $\mathbf{v}$, we choose it to be one of
the basis functions $\boldsymbol{\varphi}_{ij}$. For $i = 1,
\ldots, n_d$ and $j = 1, \ldots, n_g$ this results in the system
of exactly $n_d \cdot n_g$ linear equations
\begin{equation}
 (- \mathbf{\hat{S}}_1 + \mathbf{\hat{S}}_2)  \mathbf{a} + \beta ( \mathbf{\hat{S}}_3 +  \mathbf{\hat{S}}_5) \mathbf{a} + \beta^2 \mathbf{\hat{S}}_4  \mathbf{a} = 0,
\label{FEM EVP}
\end{equation}
where $ \mathbf{a}^\intercal = (a_{ij}) = ([a_{11}, \ldots,
a_{n_d1}], [a_{12}, \ldots, a_{n_d2}], \ldots, [a_{1n_g}, \ldots,
a_{n_dn_g}])$ (the subscript $_{ij}$ here refers to the
$j^{\text{th}}$ element of the $i^{\text{th}}$ subvector).
Since for any square matrix $\mathbf{M}$ of dimension $n_d n_g
\times n_d n_g$
\begin{equation}
\begin{array}{c}
\boldsymbol{\varphi}_{pm}^\intercal \mathbf{M}
\boldsymbol{\varphi}_{ij} = \hat{\varphi}_m(y) \cdot M_{pi}(y)
\cdot \hat{\varphi}_j
\end{array}
\end{equation}
holds, the matrices $\mathbf{S}$ turn to be of the following form
\begin{equation}
\begin{array}{l}
 \mathbf{\hat{S}}_1(pm, ij) = \int \hat{\varphi}_m S_{1pi} \hat{\varphi}_j \,dy, \\
 \mathbf{\hat{S}}_2(pm, ij) = \int (\hat{\varphi}_m)' S_{2pi} (\hat{\varphi}_j)' \,dy, \\
 \mathbf{\hat{S}}_3(pm, ij) = \int (\hat{\varphi}_m)' S_{3pi} \hat{\varphi}_j \,dy, \\
 \mathbf{\hat{S}}_4(pm, ij) = \int \hat{\varphi}_m S_{2pi} \hat{\varphi}_j \,dy, \\
 \mathbf{\hat{S}}_5(pm, ij) = \int \hat{\varphi}_m S_{3pi} (\hat{\varphi}_j)' \,dy,
\end{array}
\end{equation}
where the indices $pm$ and $ij$ have the same meaning as in the
definition of the vector $\mathbf{a}$.

The solution of the quadratic eigenvalue problem (\ref{FEM EVP})
with $\beta$ as an eigenvalue can be found by introducing an auxiliary vector $\mathbf{b}=\beta \mathbf{a}$. (\ref{FEM EVP}) can then be transformed into the following linear eigenvalue problem
\begin{equation}
 \left(%
\begin{array}{cc}
  \mathbf{0} & \mathbf{1} \\
  - \mathbf{\hat{S}}_4^{-1} (-\mathbf{\hat{S}}_1 + \mathbf{\hat{S}}_2) & - \mathbf{\hat{S}}_4^{-1} (\mathbf{\hat{S}}_3 + \mathbf{\hat{S}}_5) \\
\end{array}%
\right)
\left(%
\begin{array}{c}
  \mathbf{a} \\
  \mathbf{b} \\
\end{array}%
\right) = \beta \left(%
\begin{array}{c}
  \mathbf{a} \\
  \mathbf{b} \\
\end{array}%
\right). \label{Transformed Quadratic EVP}
\end{equation}
\noindent This is a quite straightforward, but expensive approach,
as the dimension of the transformed problem is doubled in
comparison to the original one. Other more involved approaches to
tackle a quadratic eigenvalue problem can be found e.g.\ in
\cite{TiM01}. We apply standard general eigenvalue solvers as
embedded within the LAPACK \cite{LAPACK} package.
Specialized solvers could be employed, provided that an initial
guess for the propagation constant, or a range of possible
eigenvalues, are available for the problem at hand. On the other
hand, there are situations where all the propagation constants
$\beta$ and corresponding functions $\mathbf{u}$ need to be found
together, e.g.\ if one wants to expand a 3D field in terms of
vectorial modes of some channel waveguide, as required for the
implementation of transparent boundary conditions
\cite{ISK09}.

While the entire 2D problem could also be solved directly by means
of a Finite Element method, the number of degrees of freedom in
such cases would be much higher than when solving the 1D equations
(\ref{vms:System For YEx and YHx}) using the Finite
Element Method; instead of having to use a triangulation of the entire 2D domain,
only 1D finite elements are needed; furthermore, the number of degrees
of freedom on each node is equal to the number of modes in the expansion,
which is typically a small number.

\subsection{Piecewise constant refractive index distribution}
\label{sec:Waveguides with piecewise constant refractive index
distribution}

If a waveguide has a piecewise constant rectangular refractive
index profile, it can be divided by vertical lines into slices
with constant refractive index distribution along the
$y$-direction. In each of these slices the matrices $\mathbf{S}$
do not depend on $y$. Then (\ref{vms:System For YEx and YHx}) can
be rewritten in a more familiar manner: Inside each of the slices
$\mathbf{u}$ should satisfy a system of second order differential
equations with constant coefficients $\mathbf{S}$ and a parameter
$\beta^2$
\begin{equation}
\mathbf{S}_1 \mathbf{u} + \mathbf{S}_2 \mathbf{u}'' = \beta^2
\mathbf{S}_2 \mathbf{u}, \label{System For YEx and YHx: pc
waveguide}
\end{equation}
together with the continuity conditions
(\ref{ContinuityConditionsAcrossTheInterfaces}). Moreover the
matrices $\mathbf{S}_1$ and $\mathbf{S}_2$ are block-diagonal in
such a way that the equations for the functions $\mathbf{Y}^{E_x}$
and $\mathbf{Y}^{H_x}$ decouple inside each of the slices;
coupling occurs only across the vertical interfaces.

Inside each slice a particular solution of the system (\ref{System
For YEx and YHx: pc waveguide}) can be readily written as
\begin{equation}
\mathbf{u} = c \nexp{\alpha y} {\mathbf{p}}
\label{GeneralSolutionOfReducedSystem}
\end{equation}
\noindent with some constants $c, \alpha$ and a vector
$\mathbf{p}$. By substituting
(\ref{GeneralSolutionOfReducedSystem}) into (\ref{System For YEx
and YHx: pc waveguide}) we find a generalized eigenvalue problem
with $\eta^2 = \beta^2 - \alpha^2$ as an eigenvalue:
\begin{equation}
\mathbf{S}_1 \mathbf{p} = \eta^2 \mathbf{S}_2 \mathbf{p}.
\label{EquationForAlphaAndVectorP}
\end{equation}

So inside each of the slices with uniform permittivity along the
$y$-axis the function $\mathbf{u}$ can be represented as
\begin{equation}
\mathbf{u} = \sum \limits_j \left( c_{1j} \nexp{\sqrt{\beta^2 -
\eta_j^2} y} + c_{2j} \nexp{- \sqrt{\beta^2 - \eta_j^2} y} \right)
\mathbf{p}_j \label{u in each of the slices}
\end{equation}
with eigenvalues $\eta_j$ and corresponding eigenvectors
$\mathbf{p}_j$ from (\ref{EquationForAlphaAndVectorP}).

By matching the solutions of the each individual slab across the
vertical interfaces using
(\ref{ContinuityConditionsAcrossTheInterfaces}) and looking only
for exponentially decaying solutions for $y \rightarrow \pm
\infty$, one can obtain an eigenvalue problem
\begin{equation}
 \mathbf{M}(\beta) \mathbf{c} = 0.
\label{EVP for pc-waveguide}
\end{equation}
The vector $\mathbf{c}$ consists of all unknown coefficients
$c_{1j}$ and $c_{2j}$ from the representations of $\mathbf{u}$
(\ref{u in each of the slices}) on all individual slices. The
matrix $\mathbf{M}$ depends on $\beta$ in a non-linear, even
non-polynomial way. One of the strategies to tackle this is at
first to specify a range of admissible values $\beta \in [I_1,
I_2]$, where solutions $\beta$ are sought. As we are looking only
for propagating modes, with decaying field (\ref{u in each of the
slices}) at $y \rightarrow \pm \infty$, $I_1$ should be not
smaller than the biggest eigenvalue $\eta_j$ of
(\ref{EquationForAlphaAndVectorP}) in the left-most and the
right-most slabs. At the same time we require that there exists at
least one oscillating function in at least one vertical slab. So
$I_2$ should be smaller than the biggest eigenvalue $\eta_j$ of
(\ref{EquationForAlphaAndVectorP}) of all the constituting slabs,
except the left- and the right-most ones. Once this interval is at
hand, we scan through it looking for a $\beta$ such that the
matrix $\mathbf{M}(\beta)$ has at least one zero eigenvalue.
Obviously, to find a non-trivial solution with certain accuracy
requires some iterations. Moreover a large step size might lead to
missing some roots while scanning the interval.

Once a nontrivial solution $\beta$, $\mathbf{c}$ of (\ref{EVP for
pc-waveguide}) is at hand, $\mathbf{u}$ can be reconstructed using
(\ref{u in each of the slices}). And then all field components can
be obtained according to expressions (\ref{vms:Components Ey, Ez,
Hy and Hz}) together with (\ref{5CA}) or (\ref{3CA}).

\section{Relation with the Effective Index Method}
\label{sec:Connection to the Effective Index Method}

In the following section we are going to show what happens if only
a single, TE or TM, slab mode is taken into account in VEIM5
(\ref{5CA}).
Using the variational reasoning we will rigorously derive an
analog to the Effective Index method.

\subsection{TE polarization}
\label{sec:TE Polarization} Let us take only one TE slab mode with
propagation constant $\beta_{\text{r}}$ from a reference slice
$\text{r}$ with permittivity distribution
$\varepsilon_{\text{r}}(x)$, and use it to represent the vectorial
field profile of the complete waveguide as in eqn.\ (\ref{5CA}).
Due to the fact that $X^{E_x} \equiv 0$, according to
(\ref{vms:System For YEx and YHx}) the unknown function $Y^{H_x}$
satisfies the eqn.\
\begin{equation}
\begin{array}{r}
\displaystyle A_{41} Y^{H_x} + \big(- \iu A_{42} ( A_{21} + A_{22}
A_{51}^{-1} A_{52} )^{-1} A_{23} (Y^{H_x})' \big)' = \\
\displaystyle \beta^2 \big(- \iu A_{42} ( A_{21} + A_{22}
A_{51}^{-1} A_{52} )^{-1} A_{23} \big) Y^{H_x}.
\end{array}
\end{equation}

After some manipulations, using the relations between the modal
components $\chi^{E_y}$, $\chi^{H_x}$ and $\chi^{H_z}$ of the slab
mode the above relation can be rewritten as follows
\begin{equation}
\Big(\frac{1}{\varepsilon_{\text{eff}}}  (Y^{H_x})' \Big)' + k^2
Y^{H_x} = \beta^2 \frac{1}{\varepsilon_{\text{eff}}} Y^{H_x}
\label{vms:EIM with TE mode}
\end{equation}
\noindent with
\begin{equation}
\varepsilon_{\text{eff}}(y) = \frac{\beta^2_{\text{r}}}{k^2} +
\frac{\langle \chi^{E_y}, (\varepsilon(x,y) -
\varepsilon_{\text{r}}(y)) \chi^{E_y}\rangle}{\langle \chi^{E_y},
\chi^{E_y} \rangle}. \label{New Effective Index with TE mode}
\end{equation}

This looks exactly as a TM mode equation, similar to the standard
Effective Index Method.
In the reference slice one has $\varepsilon =
\varepsilon_{\text{r}}$, and the effective permittivity
$\varepsilon_{\text{eff}}$ is equal to the squared effective index
of the mode of the reference slice $\beta^2_{\text{r}} / k^2$. In
other slices this squared effective index is modified by the
difference between the local permittivity and that of the
reference slice, weighted by the local intensity of the
fundamental component of the reference mode profile. Hence, on the
contrary to the EIM, even in slices where no guided mode exist,
the effective permittivity can still be rigorously defined.

Now it is instructive to see how the mode profile adjusts both in
the reference slabs and elsewhere. Inside a slice with constant
permittivity $\varepsilon_{\text{eff}}$, eqn.\ (\ref{vms:EIM with
TE mode}) permits solutions of the form
\begin{equation}
Y^{H_x} =  c_{+} \nexp{\iu \alpha y} + c_{-} \nexp{- \iu \alpha y}
\label{General Solution of the EIM with TE mode}
\end{equation}
for arbitrary constants $c_{+}$ and $c_{-}$ and with
\begin{equation}
\beta^2 + \alpha^2 = k^2 \varepsilon_{\text{eff}}. \label{EIM TE
Rotation}
\end{equation}
With the abbreviation $\rho^2 = k^2 \varepsilon_{\text{eff}}$ from
(\ref{vms:Components Ey, Ez, Hy and Hz}) it follows that
\begin{equation}
\begin{array}{l}
\displaystyle Y^{H_z} = \frac{\beta_{\text{r}} \beta}{\rho^2}
\big(c_{+}
\nexp{\iu \alpha y} + c_{-} \nexp{- \iu \alpha y} \big), \\
\vspace{0.15cm} \displaystyle Y^{E_y} = Y^{H_z}, \\
\vspace{0.15cm} \displaystyle Y^{E_z} = \frac{\beta_{\text{r}}
\alpha}{\rho^2} \big(c_{+} \nexp{\iu \alpha y} - c_{-} \nexp{- \iu
\alpha y} \big), \\ \vspace{0.15cm} \displaystyle Y^{H_y} = -
Y^{E_z}.
\end{array}
\end{equation}
By introducing an angle $\theta$ such that $\cos \theta = \beta /
\rho$, one can write
\begin{equation}
\begin{array}{r}
\displaystyle \Big(
\begin{array}{c}
  Y^{E_x}, Y^{E_y}, Y^{E_z} \\
  Y^{H_x}, Y^{H_y}, Y^{H_z}
\end{array}
\Big)(y) = c_{+} \frac{\beta_{\text{r}}}{\rho} \nexp{\iu \rho \sin
\theta y} \Big(
\begin{array}{lrr}
   0, &  \cos \theta, & \sin \theta \\
   \rho / \beta_{\text{r}}, & -\sin \theta, & \cos \theta
\end{array}
\Big) + \\
\displaystyle + c_{-} \frac{\beta_{\text{r}}}{\rho} \nexp{- \iu
\rho \sin \theta y} \Big(
\begin{array}{lrr}
   0, &  \cos \theta, & -\sin \theta \\
   \rho / \beta_{\text{r}}, & \sin \theta, & \cos \theta
\end{array}
\Big).
\end{array}
\label{vms:EIM rotated TE slab mode}
\end{equation}

If we use the principal square roots of $\alpha^2$ and $\rho^2$
for $\alpha$ and $\rho$, and the principal inverse cosine for
$\theta$ eq.\ \ref{vms:EIM rotated TE slab mode} can be
interpreted as follows. In the slice where the reference slab mode
lives $\rho = \beta_{\text{r}}$, and we find that functions $Y$
act as a rotation of the slab mode, such that the projection of
the propagation constant of this mode onto the $z$-axis will match
the global propagation constant $\beta$. In other slices, in
addition to the rotation of the $y$ and $z$ components of the slab
mode, the $x$ component is scaled by $\rho / \beta_{\text{r}}$.

\subsection{TM polarization}
\label{sec:TM Polarization}

Analogously, the eqn.\ (\ref{vms:System For YEx and YHx}) can be
rewritten for a single TM mode, with a field template as in
(\ref{5CA}). We now have $X^{H_x} \equiv 0$ in eqn.\
(\ref{vms:System For YEx and YHx}) and using the properties of the
TM slab mode, the original equation for the unknown function
$Y^{E_x}$,
\begin{equation}
\begin{array}{r}
\displaystyle A_{11} Y^{E_x} + \big(- \iu A_{12} \big( A_{51} +
A_{52}
A_{21}^{-1} A_{22} \big)^{-1} A_{53} (Y^{E_x})' \big)' =
\\
\displaystyle \beta^2 \big(- \iu A_{12} \big( A_{51} + A_{52}
A_{21}^{-1} A_{22} \big)^{-1} A_{53} \big) Y^{E_x},
\end{array}
\end{equation}
can be rewritten as
\begin{equation}
\Big(\frac{1}{\varepsilon_{1\text{eff}}}  (Y^{E_x})' \Big)' + k^2
\varepsilon_{2} Y^{E_x} = \beta^2
\frac{1}{\varepsilon_{1\text{eff}}} Y^{E_x}, \label{vms:EIM with
TM mode}
\end{equation}
\noindent with
\begin{equation}
\displaystyle
\begin{array}{l}
\displaystyle \varepsilon_{1\text{eff}}(y) =
\frac{\beta_{\text{r}}^2}{k^2} \frac{\langle \chi^{E_z},
\varepsilon_{\text{r}}(x) \chi^{E_z} \rangle}{\langle \chi^{E_z},
\varepsilon(x,y) \chi^{E_z} \rangle} + \frac{\langle \chi^{H_y},
\chi^{H_y} \rangle}{\langle \chi^{H_y},
\frac{1}{\varepsilon_{\text{r}}(x)} \chi^{H_y} \rangle}
\frac{\langle \chi^{E_z}, (\varepsilon(x,y) - \varepsilon_{\text{r}}(x)) \chi^{E_z} \rangle}{\langle \chi^{E_z}, \varepsilon(x,y) \chi^{E_z} \rangle}, \\
\displaystyle \varepsilon_{2}(y) = \frac{\langle \chi^{E_x},
\varepsilon(x,y) \chi^{E_x} \rangle}{\langle \chi^{E_x},
\varepsilon_{\text{r}}(x) \chi^{E_x} \rangle}. \label{New
Effective Index with TM mode}
\end{array}
\end{equation}

This appears to be neither a standard TE nor a TM mode equation,
but something in between, with the local refractive index
distribution appearing both in the terms with and without
derivative. In the reference slice with $\varepsilon =
\varepsilon_{\text{r}}$, the effective permittivity
$\varepsilon_{1\text{eff}}$ is equal to the squared effective
index $\beta^2_{\text{r}} / k^2$ of the mode of the reference
slice and $\varepsilon_{2} = 1$. Contrary to the EIM, even in
slices where no guided mode exists quantities that act like
effective indices can still be rigorously defined.

What concerns the mode profile, in intervals along the $y$-axis
with constant $\varepsilon_{1\text{eff}}$ and $\varepsilon_{2}$,
local solutions of eqn.\ (\ref{vms:EIM with TM mode}) are of the
form
\begin{equation}
Y^{E_x} = c_{+} \nexp{\iu \alpha y} + c_{-} \nexp{- \iu \alpha y}
\label{General Solution of the EIM with TM mode}
\end{equation}
with
\begin{equation}
\beta^2 + \alpha^2 = k^2 \varepsilon_{1\text{eff}}
\varepsilon_{2}. 
\end{equation}
\noindent Let us denote the right hand side of eqn.\ (\ref{EIM TE
Rotation}) as $\rho^2 = k^2 \varepsilon_{1\text{eff}}
\varepsilon_{2}$ and $\varepsilon_{3}(y) = \frac{\langle
\chi^{E_z}, \varepsilon(x,y) \chi^{E_z} \rangle}{\langle
\chi^{E_z}, \varepsilon_{\text{r}}(x) \chi^{E_z} \rangle}$, then
according to eqn.\ (\ref{vms:Components Ey, Ez, Hy and Hz}) one
obtains
\begin{equation}
\begin{array}{l}
\displaystyle Y^{H_z} = \frac{\beta_{\text{r}} \alpha
\varepsilon_{2}}{\rho^2} \big(c_{+} \nexp{\iu \alpha y}
- c_{-} \nexp{- \iu \alpha y} \big), \\
\vspace{0.15cm} \displaystyle Y^{E_y} = -
\frac{1}{\varepsilon_{3}}
Y^{H_z}, \\
\vspace{0.15cm} \displaystyle Y^{H_y} = \frac{\beta_{\text{r}}
\beta \varepsilon_{2}}{\rho^2}
\big(c_{+} \nexp{\iu \alpha y} + c_{-} \nexp{- \iu \alpha y} \big), \\
\vspace{0.15cm} \displaystyle Y^{E_z} = \frac{1}{\varepsilon_{3}}
Y^{H_y}.
\end{array}
\end{equation}
\noindent By introducing an angle $\theta$ such that $\cos \theta
= \beta / \rho$, one can write
\begin{equation}
\begin{array}{r}
\displaystyle \Big(
\begin{array}{c}
  Y^{E_x}, Y^{E_y}, Y^{E_z} \\
  Y^{H_x}, Y^{H_y}, Y^{H_z}
\end{array}
\Big)(y) = c_{+} \frac{\beta_{\text{r}}\varepsilon_{2}}{\rho}
\nexp{\iu \rho \sin \theta y} \Big(
\begin{array}{lrr}
   \rho / \beta_{\text{r}}\varepsilon_{2}, &  - \varepsilon_{3}^{-1} \sin \theta, & \varepsilon_{3}^{-1} \cos \theta \\
   0, & \cos \theta, & \sin \theta
\end{array}
\Big) + \\
\displaystyle + c_{-} \frac{\beta_{\text{r}}
\varepsilon_{2}}{\rho} \nexp{- \iu \rho \sin \theta y} \Big(
\begin{array}{lrr}
   \rho / \beta_{\text{r}}\varepsilon_{2}, &  \varepsilon_{3}^{-1} \sin \theta, & \varepsilon_{3}^{-1} \cos \theta \\
   0, & \cos \theta, & - \sin \theta
\end{array}
\Big).
\end{array}
\label{vms:EIM rotated TM slab mode}
\end{equation}

In the reference slice $\rho = \beta_{\text{r}}$ and we find that
the functions $Y$ also in this case act as a rotation of the slab
mode. In all other slices, while the $y$- and $z$-components of
the magnetic field are just rotated by the angle $\theta$, the
electric $y$ and $z$ components are not only rotated, but also
scaled by $\varepsilon_{3}^{-1}$. In addition to this the
$x$-component is scaled by $\rho / \beta_{\text{r}}\varepsilon_{2}$.

\section{Relation with the Film Mode Matching Method}
\label{sec:Connection to the Film Mode Matching Method}

As we could see in the previous section if only one, TE or TM,
slab mode is used to expand the total field profile using the 5
component expansion (\ref{5CA}), the variational procedure leads
to functions $Y$ that act as a rotation. Then the field
representation inside the slice where the slab mode lives
replicates the field ansatz of the FMM (cf.\ \autoref{sec:Modal
Field Ansatz}). In the following we look at the the case when
multiple TE and TM slab modes appear in the 5 component expansion
(\ref{5CA}).

Let us rewrite the second, third, fifth and sixth equations of
(\ref{EquationsForYFunctions}) as
\begin{equation}
\begin{array}{l}
\displaystyle \mathbf{I}_{\text{TE}} (\mathbf{Y}^{E_y}_{\text{TE}} - \mathbf{Y}^{H_z}_{\text{TE}}) + \mathbf{I}_{\text{TM}} (\mathbf{Y}^{E_y}_{\text{TM}} + \mathbf{Y}^{H_z}_{\text{TM}}) = \mathbf{A}_{21}^{-1} \mathbf{A}_{23} ( \beta \mathbf{Y}^{H_x} - \mathbf{G}_{\text{TE}} \mathbf{Y}^{H_z}_{\text{TE}}) \\

\displaystyle - \mathbf{I}_{\text{TE}} (\mathbf{Y}^{E_y}_{\text{TE}} - \mathbf{Y}^{H_z}_{\text{TE}}) + \mathbf{I}_{\text{TM}} (\mathbf{Y}^{E_y}_{\text{TM}} + \mathbf{Y}^{H_z}_{\text{TM}}) = \mathbf{A}_{51}^{-1} \mathbf{A}_{53} ( \mathbf{G}_{\text{TM}} \mathbf{Y}^{E_y}_{\text{TM}} - \iu (\mathbf{Y}^{E_x})' ) \\

\displaystyle \mathbf{I}_{\text{TE}} (\mathbf{Y}^{E_z}_{\text{TE}} + \mathbf{Y}^{H_y}_{\text{TE}}) + \mathbf{I}_{\text{TM}} (\mathbf{Y}^{E_z}_{\text{TM}} - \mathbf{Y}^{H_y}_{\text{TM}}) = \mathbf{A}_{21}^{-1} \mathbf{A}_{23} ( \mathbf{G}_{\text{TE}} \mathbf{Y}^{H_y}_{\text{TE}} - \iu (\mathbf{Y}^{H_x})' ) \\

\displaystyle \mathbf{I}_{\text{TE}} (\mathbf{Y}^{E_z}_{\text{TE}} + \mathbf{Y}^{H_y}_{\text{TE}}) -  \mathbf{I}_{\text{TM}} (\mathbf{Y}^{E_z}_{\text{TM}} - \mathbf{Y}^{H_y}_{\text{TM}}) = \mathbf{A}_{51}^{-1} \mathbf{A}_{53} ( \beta \mathbf{Y}^{E_x} - \mathbf{G}_{\text{TM}}
\mathbf{Y^{E_z}_{\text{TM}}} )
\end{array}
\label{Rewritten Equations to show a rotation}
\end{equation}
\noindent with functions $\mathbf{Y}^a_b$ corresponding to a
vector of all the functions $Y$ related to modal component of
polarization $b$, used to expand component $a$ of the total field.
$\mathbf{G}_b$ is a diagonal matrix with propagation constants $\beta_{\text{r},j}$ of
the slab modes of polarization $b$ sitting on the diagonal.
Matrices
\begin{equation}
\begin{array}{l}
\mathbf{I_{\text{TE}}} = \left(
\begin{array}{c}
          \mathbf{1}_{ n_{\text{TE}} \times n_{\text{TE}} } \\
          \mathbf{0}_{ n_{\text{TM}} \times n_{\text{TE}} }
        \end{array} \right),
\quad \mathbf{I_{\text{TM}}} = \left(
\begin{array}{c}
          \mathbf{0}_{ n_{\text{TE}} \times n_{\text{TM}} } \\
          \mathbf{1}_{ n_{\text{TM}} \times n_{\text{TM}} }
        \end{array} \right),
\label{Matrices M}
\end{array}
\end{equation}
have been introduced to increase the readability of the equations.
Here $\mathbf{1}_d$ and $\mathbf{0}_d$
denote correspondingly the unity- and zero-matrix of a dimension
$d$, and symbols $n_{\text{TE}}$ and $n_{\text{TM}}$ denote the
number of slab modes of respectively TE and TM polarization
included in the expansion (\ref{5CA}).

Obviously, functions $\mathbf{Y}$ that satisfy
\begin{equation}
\begin{array}{ll}
 \mathbf{Y}^{E_y}_{\text{TE}} = \mathbf{Y}^{H_z}_{\text{TE}}, & \mathbf{Y}^{E_y}_{\text{TM}} = - \mathbf{Y}^{H_z}_{\text{TM}}, \\

 \beta \mathbf{Y}^{H_x} = \mathbf{G}_{\text{TE}} \mathbf{Y}^{H_z}_{\text{TE}}, \qquad & \mathbf{G}_{\text{TM}} \mathbf{Y}^{E_y}_{\text{TM}} = \iu (\mathbf{Y}^{E_x})',\\

 \mathbf{Y}^{E_z}_{\text{TE}} = - \mathbf{Y}^{H_y}_{\text{TE}}, & \mathbf{Y}^{E_z}_{\text{TM}} = \mathbf{Y}^{H_y}_{\text{TM}}, \\

 \mathbf{G}_{\text{TE}} \mathbf{Y}^{H_y}_{\text{TE}} = \iu (\mathbf{Y}^{H_x})', & \beta \mathbf{Y}^{E_x} = \mathbf{G}_{\text{TM}} \mathbf{Y}^{E_z}_{\text{TM}}
\end{array}
\label{vms:form of Y functions for FMM}
\end{equation}
\noindent are solutions of (\ref{Rewritten Equations to show a
rotation}).
Using these relations together with the first and the fourth
equations of (\ref{EquationsForYFunctions}) result in
\begin{equation}
\begin{array}{l}
 (\mathbf{Y}^{E_x})'' + (\mathbf{G}_{\text{TM}})^2 \mathbf{Y}^{E_x} = \beta^2 \mathbf{Y}^{E_x} \\
 (\mathbf{Y}^{H_x})'' + (\mathbf{G}_{\text{TE}})^2 \mathbf{Y}^{H_x} = \beta^2 \mathbf{Y}^{H_x}.
\end{array}
\label{Rewritten Equations to show a rotation 2}
\end{equation}

According to eqns.\ (\ref{Rewritten Equations to show a rotation})
and (\ref{Rewritten Equations to show a rotation 2}) all the
functions $Y$ decouple inside the slice where the slab modes
belongs to. So we can solve these equations for all the components
of $\mathbf{Y}^{E_x}$ and $\mathbf{Y}^{H_x}$ separately.
Solutions of (\ref{Rewritten Equations to show a rotation 2}) have
the form
\begin{equation}
Y^{H_x}_j = c_{+,j} \nexp{\iu \alpha_j y} + c_{-,j} \nexp{- \iu
\alpha_j y} 
\end{equation}
with
\begin{equation}
\beta^2 + \alpha_j^2 = \beta_{\text{r},j}^2. 
\end{equation}
Other components can be derived from (\ref{Rewritten Equations to
show a rotation}) as
\begin{equation}
\begin{array}{r}
\displaystyle \Big(
\begin{array}{c}
  Y^{E_x}_j, Y^{E_y}_j, Y^{E_z}_j \\
  Y^{H_x}_j, Y^{H_y}_j, Y^{H_z}_j
\end{array}
\Big)(y) = c_{+,j} \nexp{\iu \beta_{\text{r},j} \sin \theta_j y}
\Big(
\begin{array}{lrr}
   0, &  \cos \theta_j, & \sin \theta_j \\
   1, & -\sin \theta_j, & \cos \theta_j
\end{array}
\Big) + \\
\displaystyle + c_{-,j} \nexp{- \iu \beta_{\text{r},j} \sin
\theta_j y} \Big(
\begin{array}{lrr}
   0, &  \cos \theta_j, & -\sin \theta_j \\
   1, & \sin \theta_j, & \cos \theta_j
\end{array}
\Big),
\end{array}
\label{vms:FMM rotated TM slab mode}
\end{equation}
where $\cos \theta_j = \beta / \beta_{\text{r},j}$.

Hence the functions $Y_j$ corresponding to TE slab mode number $j$
rotate the original slab mode around the $x$-axis such that the
projection of its propagation constant $\beta_{\text{r},j}$ onto
the direction of propagation $z$ will be precisely the propagation
constant $\beta$ of the mode of the complete waveguide structure.
The same is true for TM slab modes.

We showed that the field ansatz of rotated slab modes, as used
locally in the Film Mode Matching method \cite{Sud93b,Bie04a} 
can be found also by the present approach where
it appears to be optimal. While in itself it might seem rather
pointless to reinvent the method, the idea behind the present
technique might be used in deriving some sort of analogue of the
FMM for full 3D scattering problems, in which the structure varies
in all 3 directions \cite{ISK09}.

\section{Numerical results}
\label{sec:Numerical Results}

We will illustrate the method with four examples. The first two
deal with waveguides with piecewise constant rectangular refractive index distribution. The third example is
a waveguide with slanted sidewall and the fourth is an indiffused
waveguide. We will use the acronym VEIM (variational effective
index method) for results of the technique as introduced in
sections \ref{sec:Variational Form of the Vectorial Mode Problem}
- \ref{sec:Connection to the Film Mode Matching Method}.

\subsection{Box-shaped waveguide}
\label{sec:Waveguide with piecewise rectangular cross-section}

\FigSideWidth{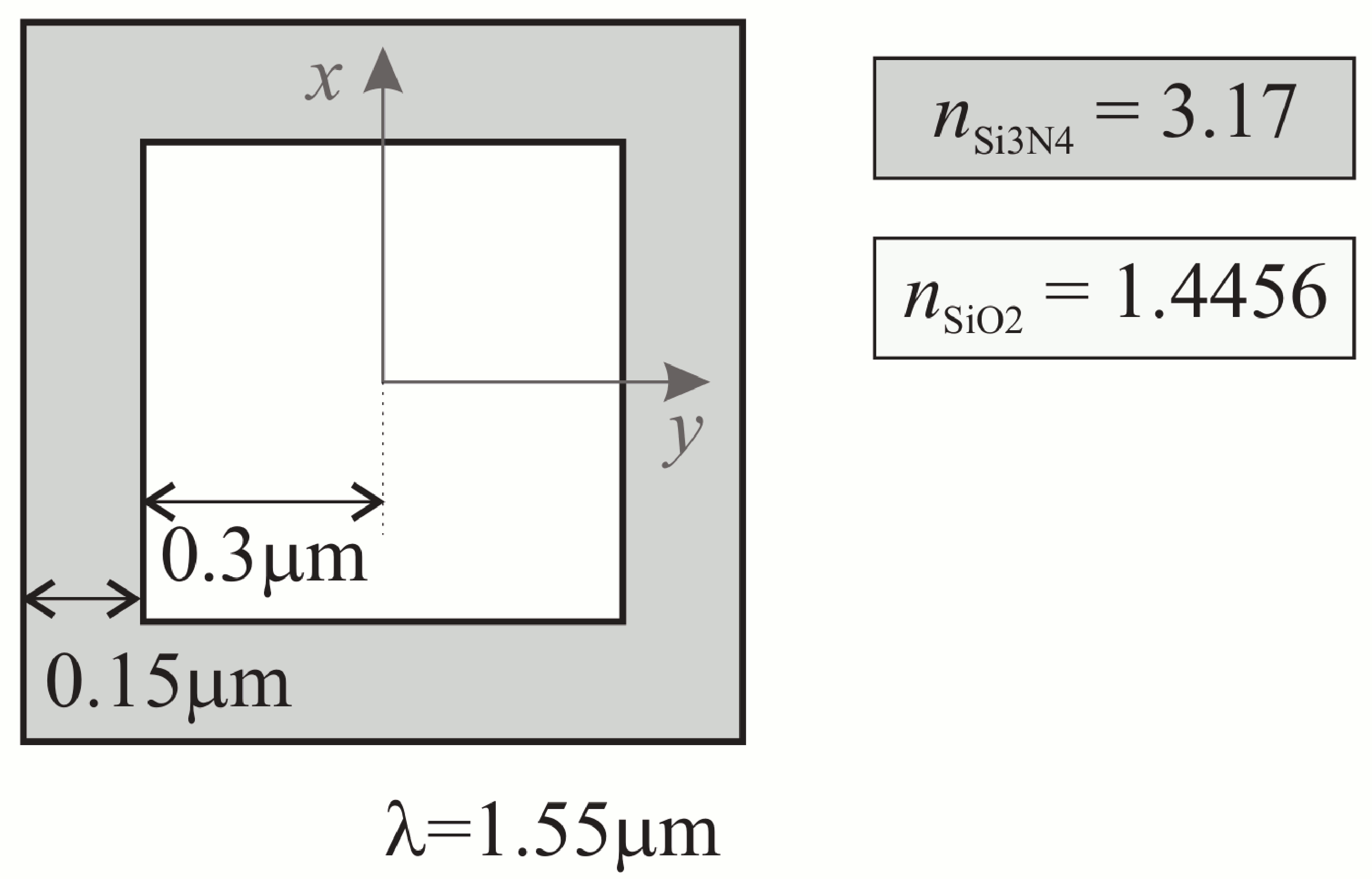}{0.35\textwidth}{Structure of
the Box-Shaped Waveguide. The vertical extents of the
computational window range from $-2.5\mu m$ to $2.5\mu m$.}

Consider the box-shaped waveguide of \reffig{OWTNMSquareWPapameters}, originating from
\cite{MMM07}. It can be divided into five vertical slices with three distinct
cross-sections (\reffig{IllustrationShellWaveguide}, left). We take slab
modes from the side walls of the box (\reffig{IllustrationShellWaveguide}, middle) to approximate the modal field
in the entire cross-section (\reffig{IllustrationShellWaveguide}, right). The waveguide will be analyzed with both
3 (Equation (\ref{3CA}) ) and 5 (Equation (\ref{5CA}) ) component approximations, denoted by VEIM3$_{a,b}$ andVEIM5$_{a,b}$,
where $a$ and $b$ are the number of TE and TM slab modes taken into account.

\FigSideWidth{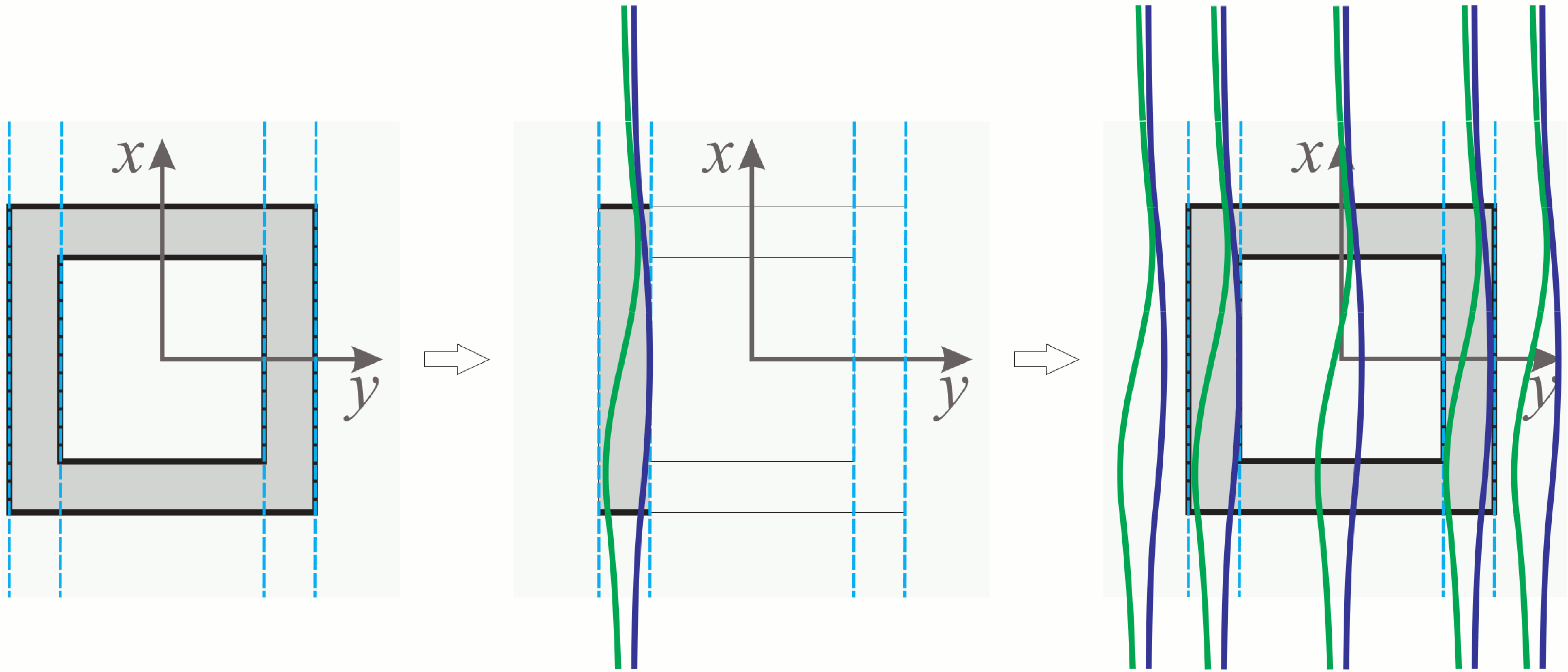}{0.6\textwidth}{Subdivision
of the waveguide into slices. Slab modes of the side walls are
used to approximate the field of the mode everywhere.}

In \reffig{CombinedBoxWg3} VEIM5$_{1,0}$ approximation of the
vectorial mode profile of the fundamental TE-like mode is shown.
In this case $\chi^{E_y}$ is multiplied by $Y^{E_y}$ and $Y^{E_z}$
to get $E_y$ and $E_z$ respectively; $\chi^{H_x}$ is multiplied by
$Y^{H_x}$ to get $H_x$; and $\chi^{H_z}$ is multiplied by
$Y^{H_y}$ and $Y^{H_z}$ to get $H_y$ and $H_z$ respectively. The
figure contains plots of all contributing functions. Consistent
with the observation in sec.\ \ref{sec:TE Polarization}, we see
that $Y^{E_y} = Y^{H_z}$ and $Y^{E_z} = - Y^{H_y}$. Note that,
contrary to the EIM, the field profile can still be visualized
even when no local guided slab mode exists.

\FIGh{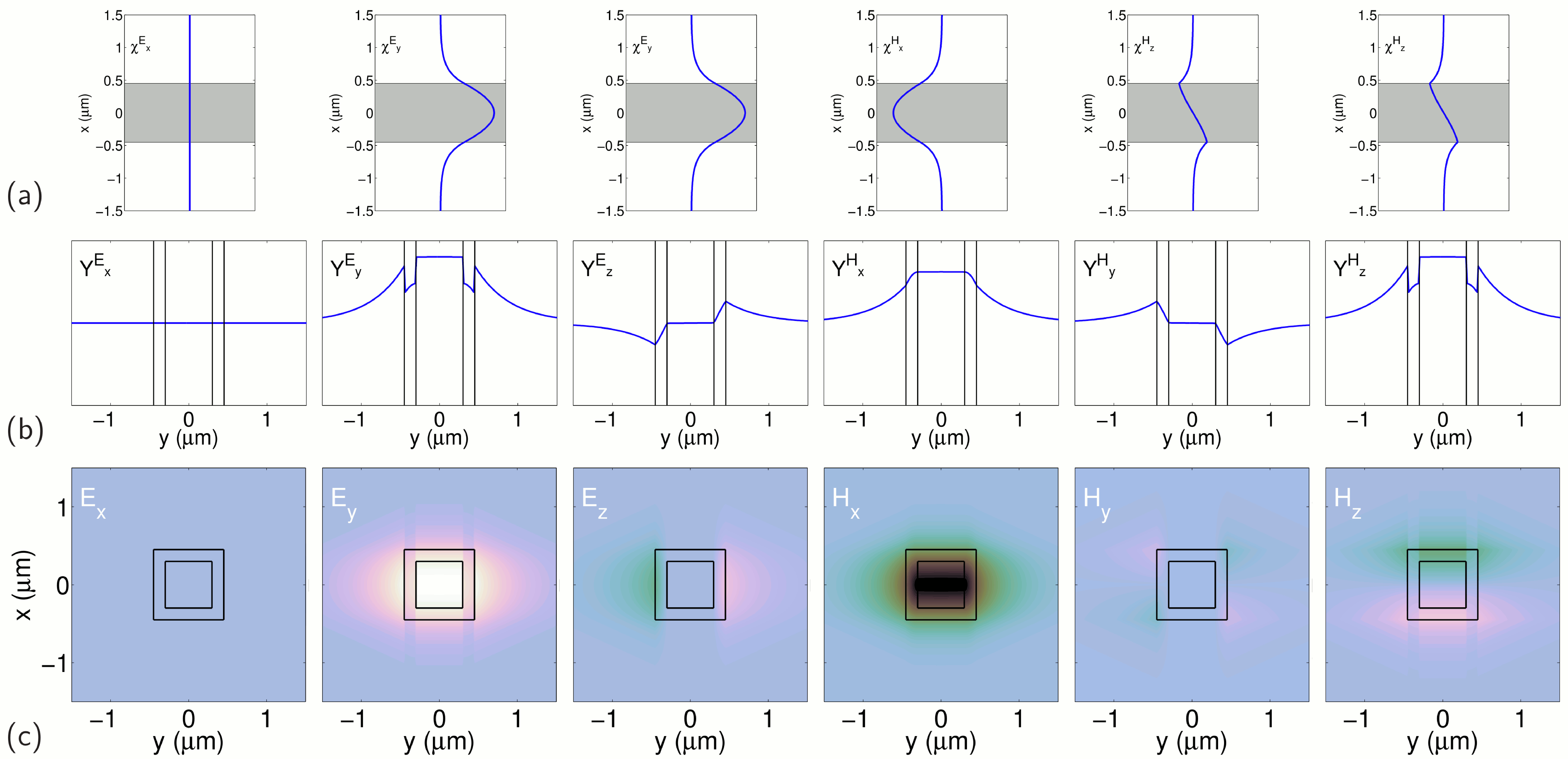}{\textwidth}{Square waveguide: (a) Functions
$\chi$ in expansion VEIM5$_{1,0}$; (b) Functions $Y$ in expansion
VEIM5$_{1,0}$; (c) Vectorial field profile VEIM5$_{1,0}$.}

Next, \reffig{CombinedBoxWg} gives an impression of the
"converged" field profile obtained using  VEIM3$_{30,30}$. The
slab mode basis has been discretized by Dirichlet boundary
conditions on the boundaries of the vertical computational window
as given in \reffig{OWTNMSquareWPapameters}. Comparison with
\reffig{CombinedBoxWg3} shows that even with a single mode in the
representation, the main features of the true field profile are
already visible. So the present method with one mode in the
expansion can very well serve as a quick tool for qualitative
analysis of the waveguide structures, while also being able to
quantitatively analyze the waveguide by using more modes in the
expansion.

\FIGh{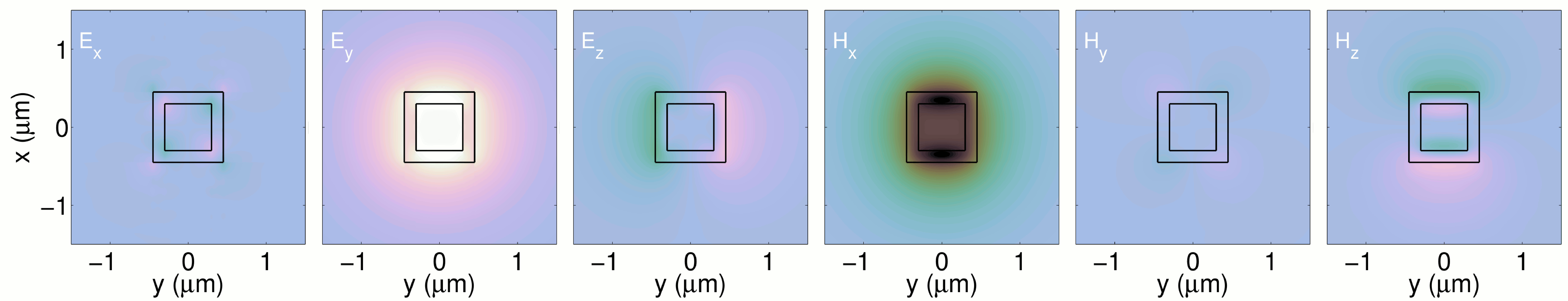}{\textwidth}{"Converged" (VEIM3$_{30,30}$)
vectorial field profiles of the fundamental TE-like mode.}

\reffig{ConvergenceBoxWCh} shows the propagation constant of the
fundamental modes of the waveguide versus the number $m$ of TE and
TM modes in the expansion VEIM3$_{m,m}$ for both the present
method and a commercial FMM solver \cite{PhoeniX}.
Both methods converge to the same value with comparable
convergence speed.

\FigSideWidth{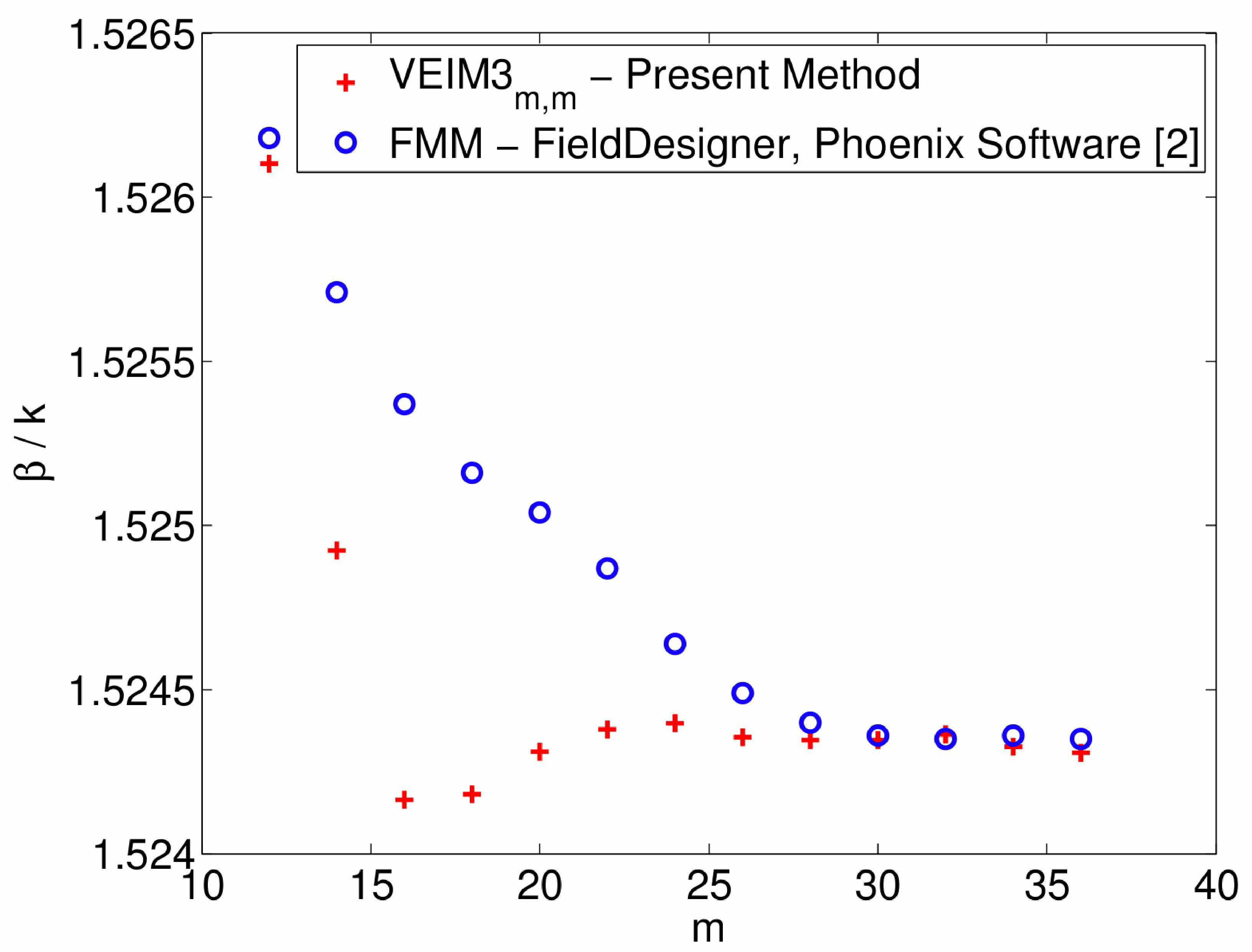}{0.4\textwidth}{Convergence of the
effective index of the fundamental TE-like mode of the box-shaped
waveguide \reffig{OWTNMSquareWPapameters}.}

\subsection{Rectangular rib waveguide}
\label{sec:Rectangular Rib Waveguide}

\FigSideWidth{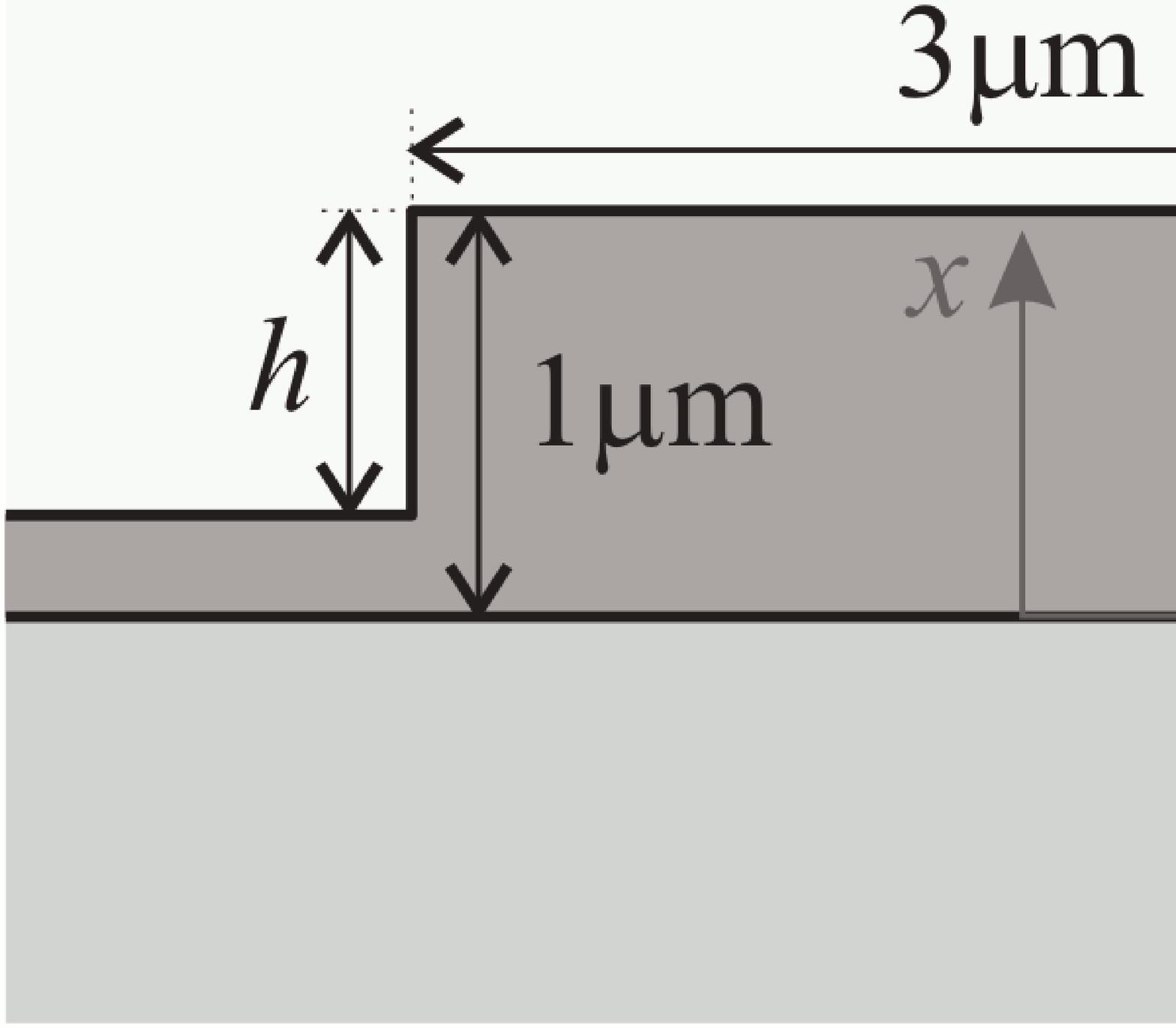}{0.4\textwidth}{Structure of
the Rib Waveguide. Vertical extents of the computational window
are $[-2, 2]\mu m$.}

In this section we consider the rib structure from
\reffig{RibWaveguideStructure}, which is used as a benchmark
waveguide in \cite{Vas97,HaS95,Loh99,IHS07}. The structure supports a
fundamental TE and TM mode for all etch depths $h$ in the range we
look in, which is [0.2, 1]. The modes are strongly polarized, and
thus it may be expected that an expansion using only TE or only TM
modes (similar to a semi-vectorial calculation) will give good
results.

At etch depths greater than $0.5\mu m$ guided modes do not exist
outside the central slice, so the EIM fails to uniquely determine
the effective refractive index of those regions. We analyze this
structure with both 3- (\ref{3CA}) and 5-component (\ref{5CA})
approximations. In the following figures we will refer to them as
VEIM3$_{a,b,c,d}$ and VEIM5$_{a,b,c,d}$ correspondingly. The
subscript letters stand for number of slab modes used in the
current approximation: $a$ -- number of TE modes from the central
slice, $b$ -- number of TE modes from the outer slice, $c$ --
number of TM modes from the central slice, $d$ -- number of TM
modes from the outer slice.

The slab modes are calculated using Dirichlet boundary conditions
on the upper and lower computational domain boundaries. Because of
this, the outer slice mode is still defined when the guided mode
of that slice goes below cut-off.

\reffig{plot_combiner_TE} and \reffig{plot_combiner_TM} show plots
of the TE and TM effective indices correspondingly using these
different expansions versus the etch depth. The figures also show
the corresponding EIM results, and, as reference, FMM results
obtained by the commercial mode solver \cite{PhoeniX}.

\FIGh{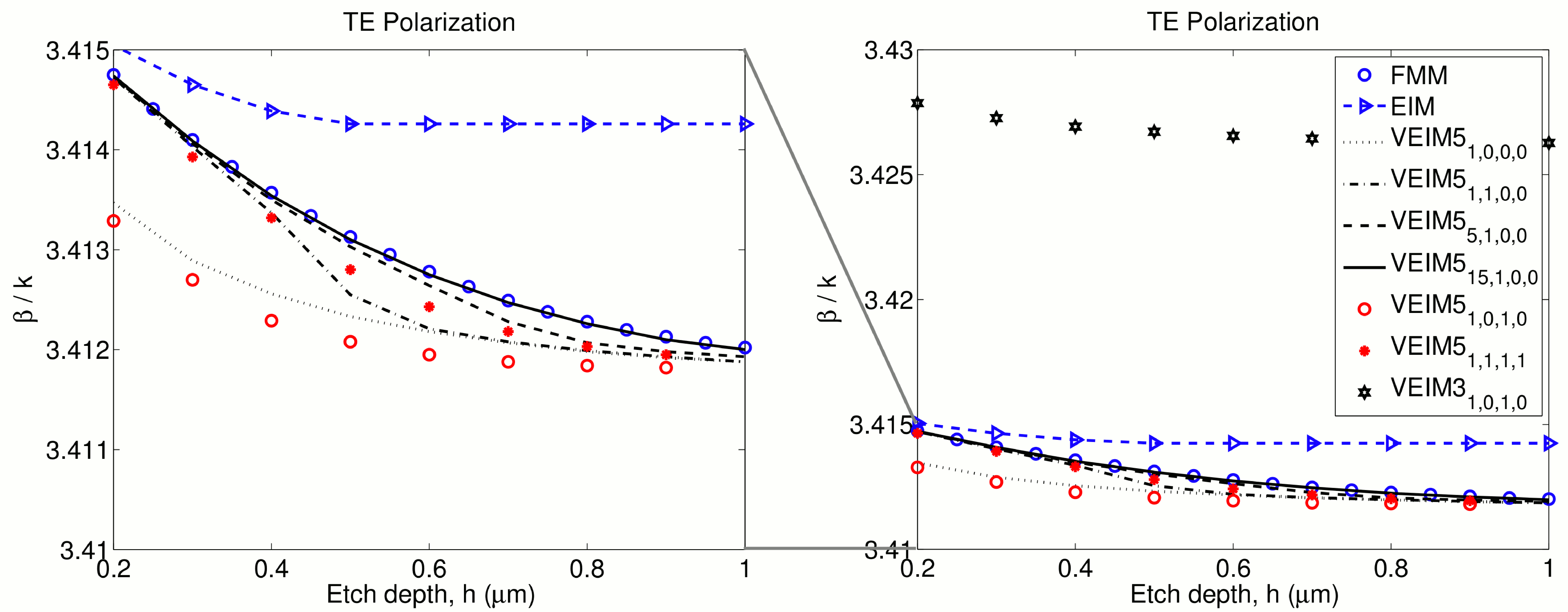}{0.9\textwidth}{Convergence of the
effective index ($\beta / k$) of the fundamental TE mode of the
rib waveguide.}

\FIGh{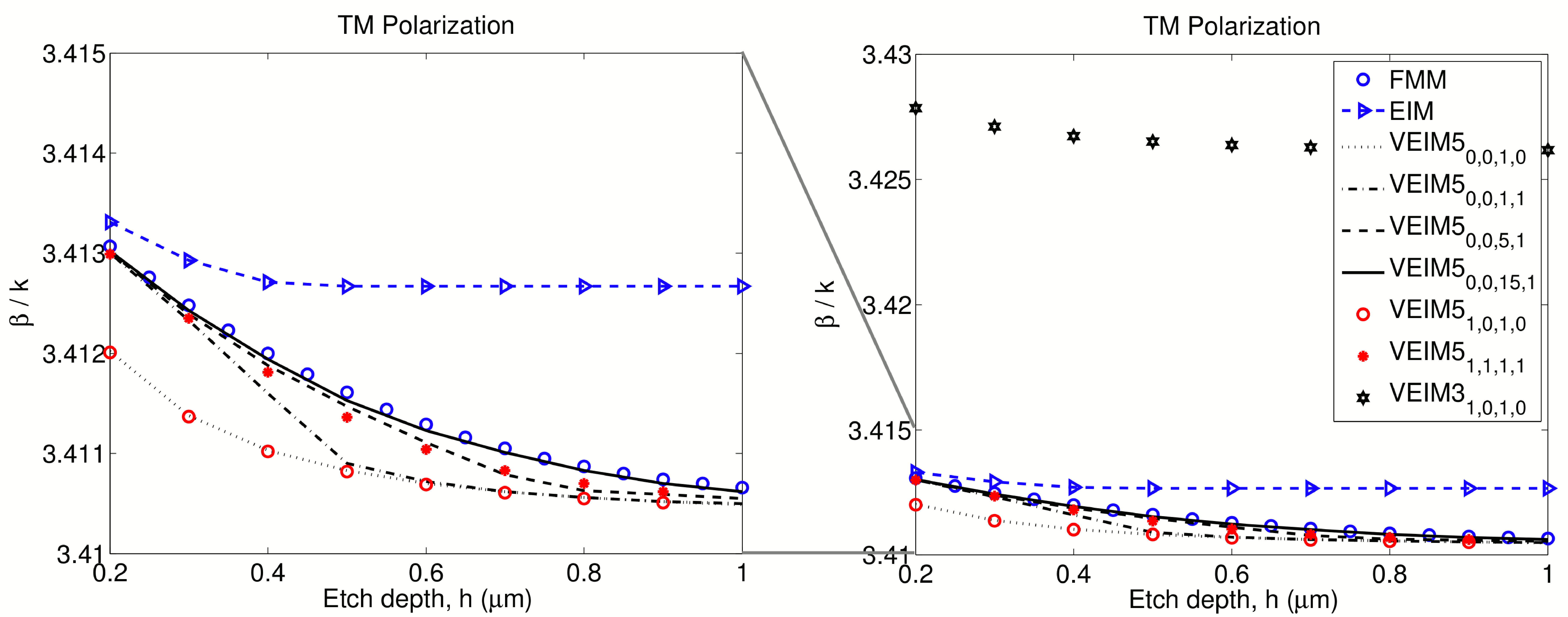}{0.9\textwidth}{Convergence of the
effective index ($\beta / k$) of the fundamental TM mode of the
rib waveguide.}

Comparing the results of our method with only one TE
(VEIM5$_{1,0,0,0}$) or TM (VEIM5$_{0,0,1,0}$) mode of the central
slice in the expansion (\ref{5CA}) to the EIM results, shows that
for larger etch depths, our results are much closer to the
reference results - especially after the outer slice has gone
below cut-off and the EIM uses the substrate refractive index as
(constant) outer slice effective index.

Adding one outer slice mode to the VEIM expansion greatly improves
its accuracy, especially if it is a guided slab mode; the
VEIM5$_{1,1,0,0}$  curves are much closer to the reference results
than the VEIM5$_{1,0,0,0}$  curves, especially at etch depths
below $0.5\mu m$.

Taking five inner and one outer slice mode VEIM5$_{5,1,0,0}$ moves
the results closer to the reference curve, while fifteen inner and
one outer slice modes VEIM5$_{15,1,0,0}$ yield results that almost
coincide with the reference. Note that these results use only TE
or only TM modes in the 5-component expansion (\ref{5CA}), i.e.\
the resulting fields are semi-vectorial; apparently a
semivectorial approximation is sufficient for an accurate
estimation of the effective indices of this structure.

The present method when using just one central slice TE
and TM mode simultaneously with the three-component-per-mode
approximation VEIM3$_{1,0,1,0}$ (\ref{3CA}) yields results that are quite far from
the reference data. Moving to the five-component-per-mode
approximation VEIM5$_{1,0,1,0}$ (\ref{5CA}), on the other hand, gives much better results.
Moreover, adding outer slice TE and TM modes VEIM5$_{1,1,1,1}$ greatly improves the
estimation of propagation constant for both, TE and TM, polarizations.

\subsection{Waveguide with non-rectangular piecewise constant cross-section}
\label{sec:Waveguide with piecewise non-rectangular cross-section}

The waveguide cross-section of \reffig{OWTNMPolConvertorStructure}
is part of a polarization rotator in InP/InGaAsP, proposed in
\cite{OSR03}. Due to its slanted sidewall, the modes of this
structure are highly hybrid.

\FigSideWidth{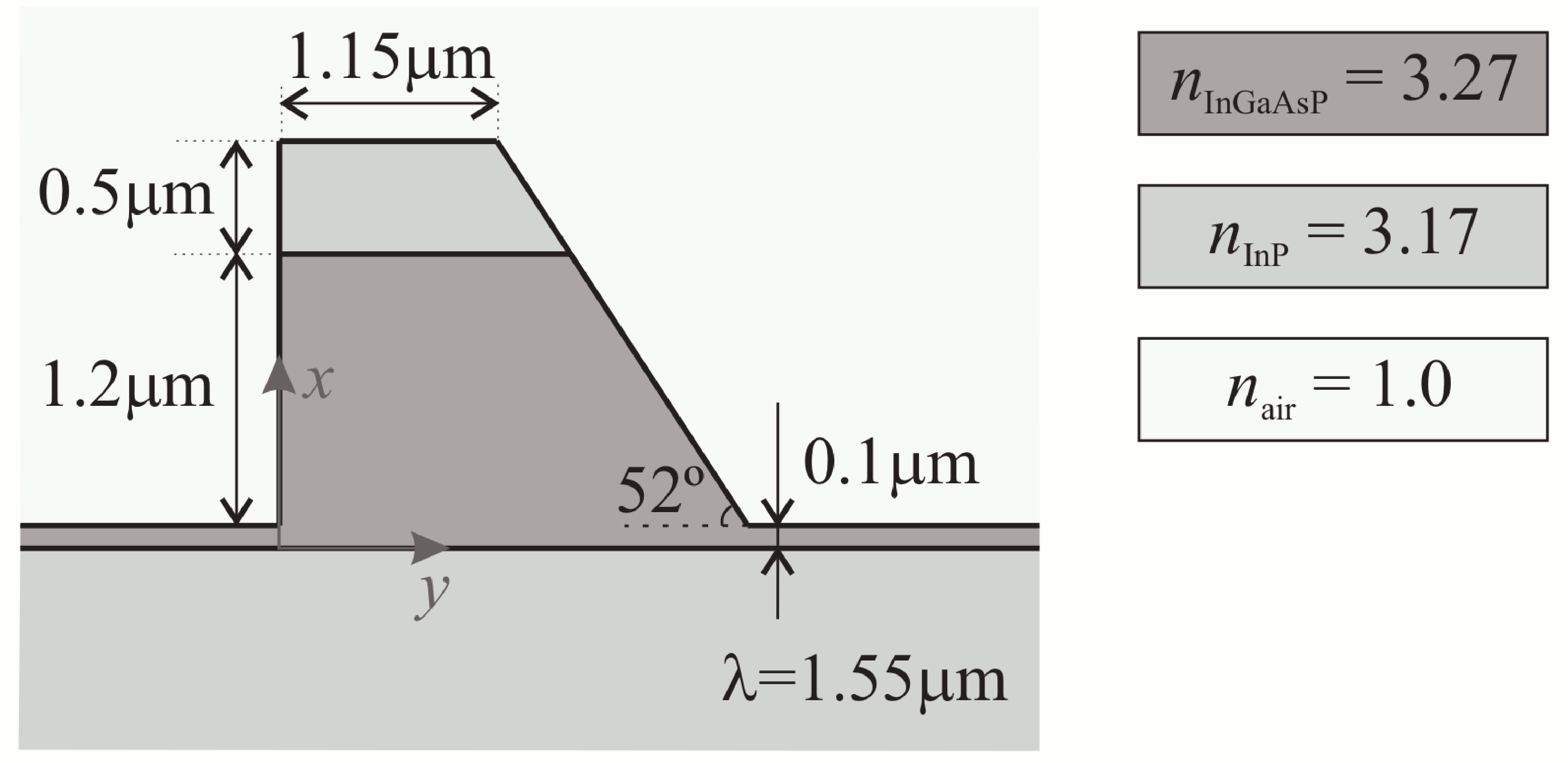}{0.4\textwidth}{Structure
of polarization converter from \cite{OSR03}. The computational
window in the calculations is $(x,y)\in[-2,2.5]\times[-2,3.5]\mu
m^2$; 50 elements are used in the finite element scheme.}

Because of the slanted sidewall, the finite element scheme is more
suitable to calculate the modes of this structure; the
semi-analytical method requires a rather large number of slices,
while the finite elements automatically take the slant into
account.

\FigSideWidth{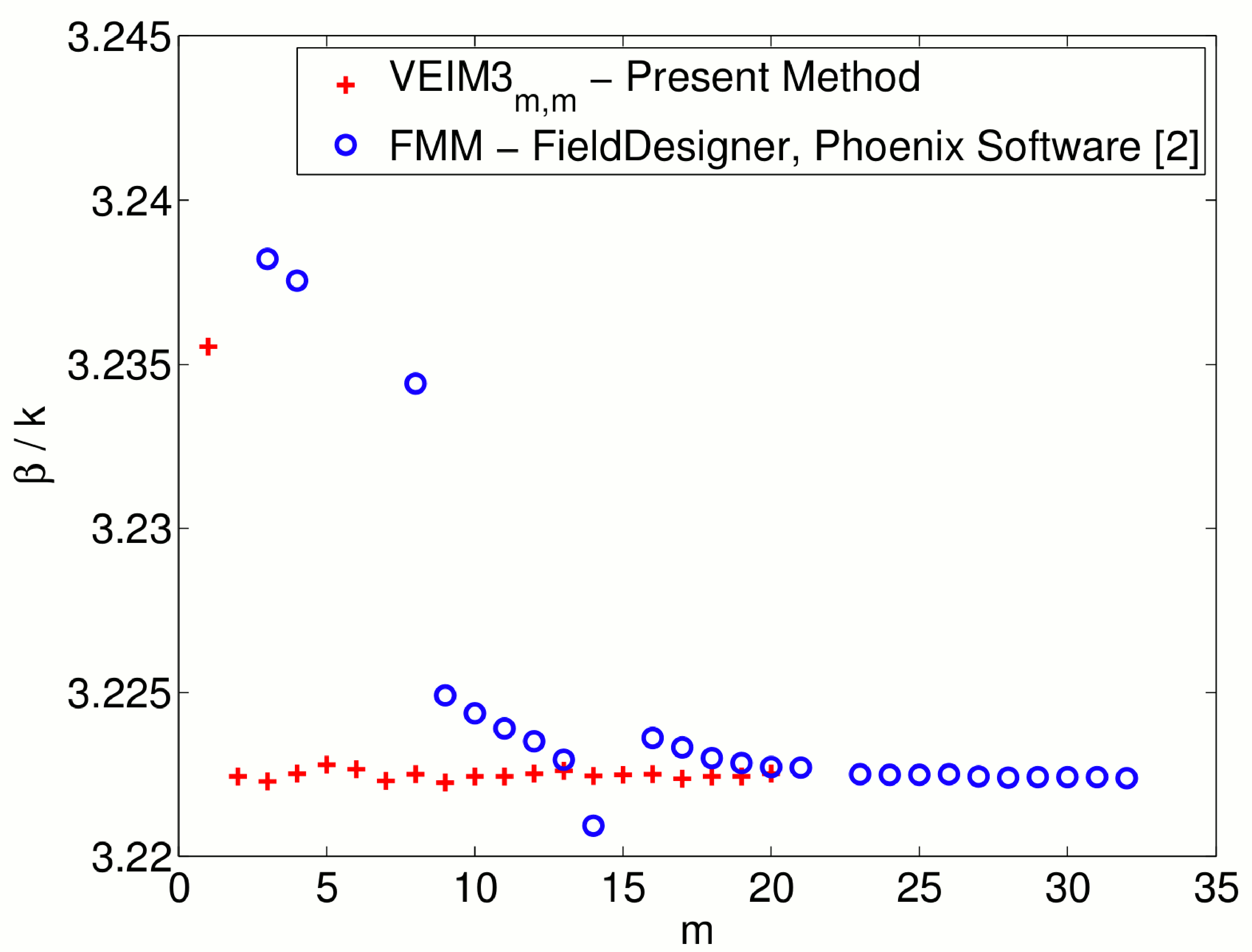}{0.4\textwidth}{Convergence of
the effective index of the fundamental mode of the polarization
converter.}

\reffig{ConvergenceSlanted} shows the convergence of the effective
index of the fundamental mode of the waveguide versus the number
of modes in the 3-component expansion VEIM3$_{a,b}$ (\ref{3CA}),
with $a$ and $b$ being numbers of TE and TM slab modes from the
central ($y \in (0, 1.15)\mu m$) slab. It also shows the
convergence of the commercial FMM mode solver
\cite{PhoeniX}, in which the structure is subdivided
into 50 slices. Remarkably, starting from just 2 TE and TM modes
in the 3-component expansion (\ref{3CA}) VEIM3$_{2,2}$, the
effective index is stable and close to the converged value of the
FMM solver; 320 modes in the FMM solver lead to an effective index
of 3.2225, while with just 7 TE and TM modes the current method
predicts already an effective index of 3.2223. The field profiles
also converge rapidly; \reffig{CombinedPolConv} shows the
vectorial fields for (a) one (VEIM3$_{1,1}$), (b) two
(VEIM3$_{2,2}$), and (c) seven (VEIM3$_{7,7}$) TE and TM modes in
the expansion (\ref{3CA}).

\FIGh{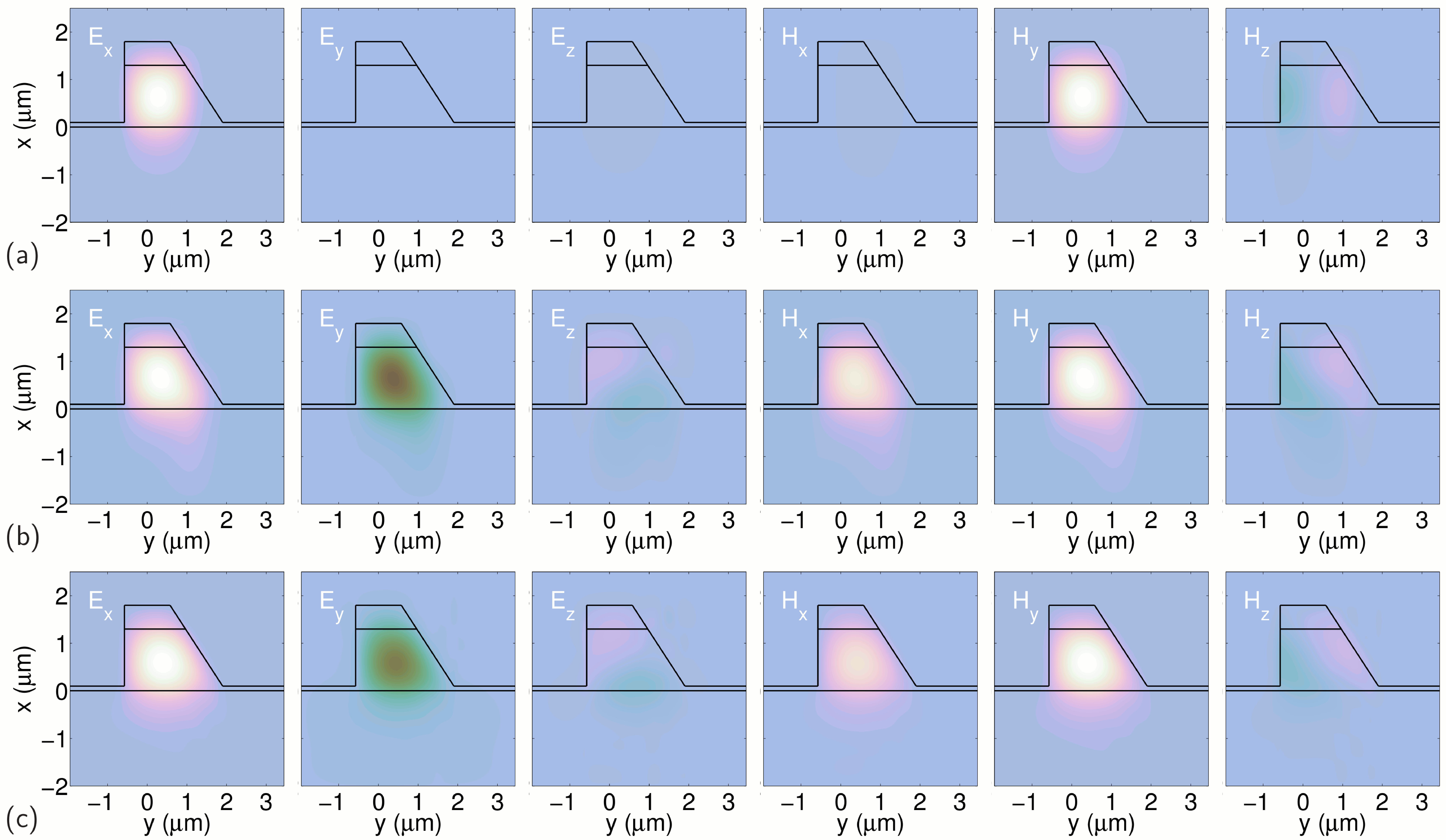}{\textwidth}{Vectorial field profile of the
fundamental mode of the polarization converter. (a) VEIM3$_{1,1}$,
(b) VEIM3$_{2,2}$, (c) VEIM3$_{7,7}$.}

\subsection{Indiffused waveguide}
\label{sec:Diffused Waveguide}

To show the flexibility of the present method we apply it to a
diffused waveguide \cite{ShB92} with a refractive index
distribution given by
\begin{equation}
n^2(x,y) = \Big\{
\begin{array}{l}
n_s^2 + n_s^2 (1.05^2 - 1) \exp(-x^2/16) \exp(-y^2/4), \ \ \ \text{if} \ \ x>0;  \\
n_c^2, \ \ \ \text{if} \ \ x<0,
\end{array}
\label{RefractiveIndDistribution of Diffused Waveguide}
\end{equation}
with $n^2_s = 2.1$, $n_c^2 = 1.0$ and $\lambda = 1.3 \mu
\text{m}$. Similar to the slanted sidewall waveguide described
above, the finite element implementation of the presented method
is the more suitable, since it takes into account the nonuniform
distribution in the y-direction of the refractive index
automatically. Vertically, the structure is subdivided into 7
layers; horizontally, 20 finite elements are used. The
computational window used in the calculations is defined as
$(x,y)\in[-1,8]\times[-6,6]\mu m^2$.

\FigSideWidth{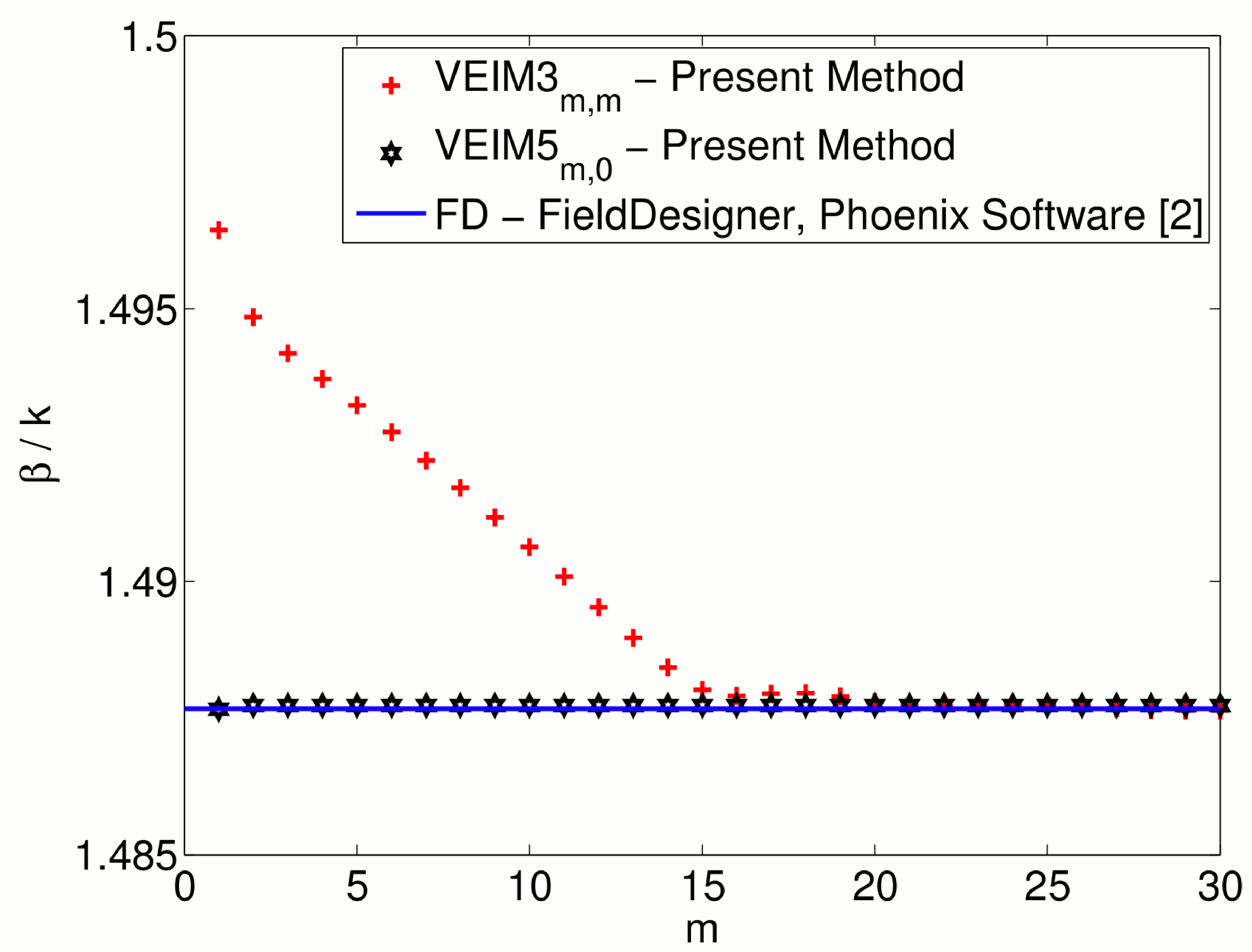}{0.4\textwidth}{Convergence of
the effective index of the fundamental mode of the diffused
waveguide.}

\reffig{ConvergenceDiffused} shows the convergence of the
effective index of the fundamental mode of the indiffused
waveguide versus the number of modes in the 3- (VEIM3$_{a,b}$) and
5-component (VEIM5$_{a,b}$) approximations, with $a$ and $b$ being
numbers of TE and TM slab modes of the central ($y=0\mu m$) slab.
The results are compared to the rigorous Finite Difference
simulation (with $129\times129$ grid points)
\cite{PhoeniX}. Since the fundamental mode is strongly
polarized, the semi-vectorial approximation appears to converge
much faster.

\FigSideWidth{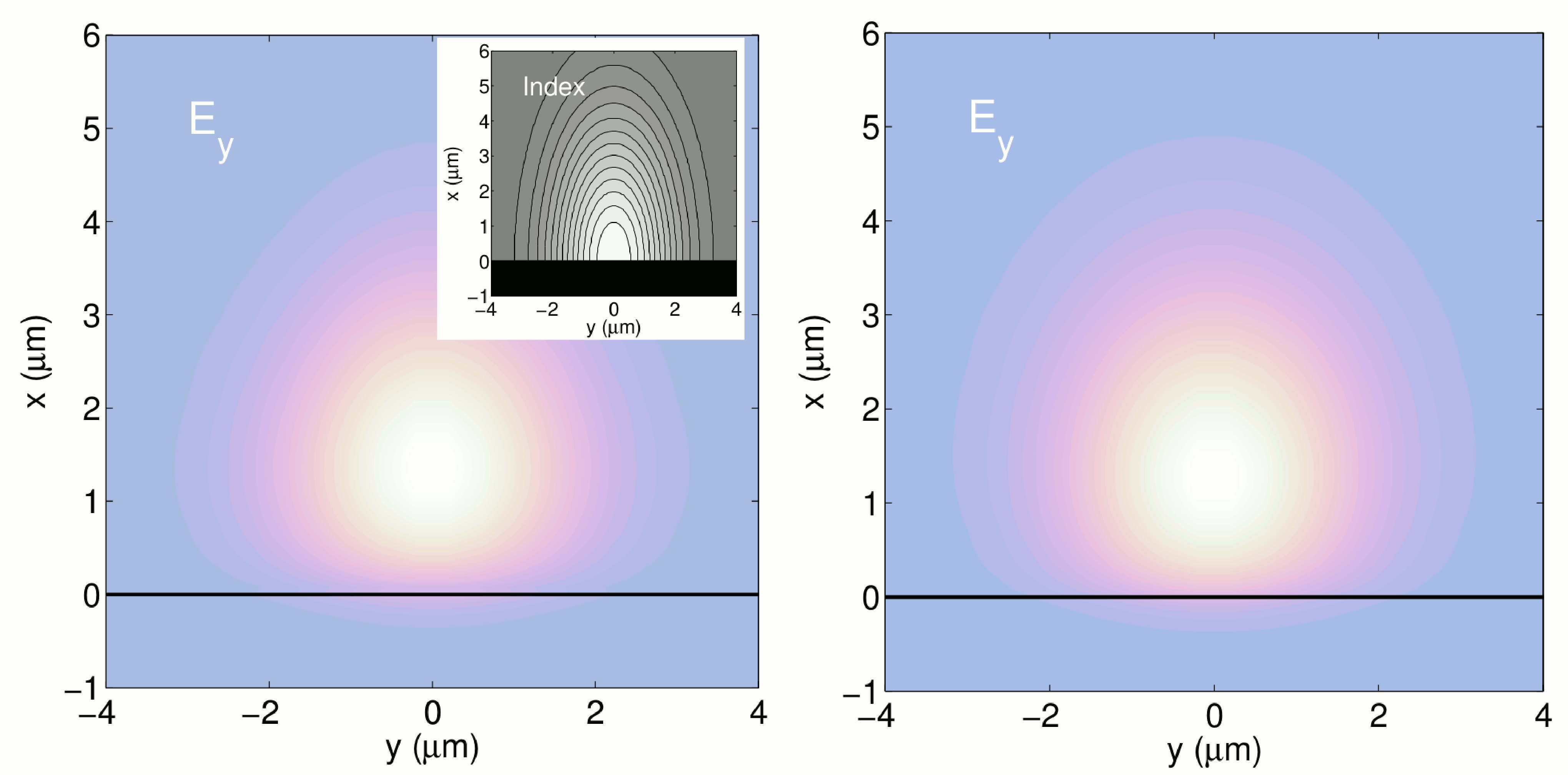}{0.5\textwidth}{Field profiles of
the dominant electric component $E_y$ of the fundamental TE mode:
left -- VEIM3$_{1,1}$, right -- VEIM3$_{15,15}$.}

On \reffig{diffused_combined2} field profiles of the dominant
electric component $E_y$ of the fundamental TE mode are shown. The
effective index $N_{\text{eff}} = \beta \lambda / 2\pi$ of the
fundamental mode on the left picture is $1.4965$ and on the right
-- $1.48802$, which compares well with the Finite Difference
simulation -- $1.48797$.

\section{Concluding remarks}
\label{sec:Concluding Remarks}

A variational method for the fully vectorial analysis of arbitrary
isotropic dielectric waveguides was developed. Similar to the
scalar approach \cite{IHS07} this method gives rather accurate
estimates of the propagation constants, sometimes even with only a
few terms in the expansion.

When applying the present method with only one slab mode in the
expansion of the modal field of the complete waveguide, this mode
is transformed in all different slices to fit the true solution
there the best. Together with the shape transformation, the
effective index of this mode is uniquely transformed. Additionally,
the expression for the transformed propagation
constant is quite simple and is certainly not more complicated,
than the calculation of a slab mode. In this way the present
procedure turns out to be a simple and still a more rigorous way to obtain a
first intuitive guess for the propagation constant and field
profile, than the standard Effective Index Method.

It turns out that in case a TE mode is used in the expansion, the
reduced equation appears to be a TM mode equation. At the same
time when a TM mode is used, the reduced equation appears to be
neither TE, nor TM mode equation, but something in between, with
the effective refractive indices appearing both under the
derivative sign and in the right part of the equation.

While in the Film Mode Matching method, rotated modes of each slice are used to locally expand the field, VEIM
uses only one set of modes everywhere. We showed that in the reference slice, where the 1D modes
are calculated, VEIM predicts exactly the same rotations as the Film Mode Matching method uses.
In the reference slice the total field profile is a superposition of these rotated 1D TE and TM modes;
in other slices, however, the components of all the 1D modes mix.

Of course the question remains - would some other combination of
slab mode components lead to faster convergence? For example, one
could imagine that in a certain case a superposition of e.g.\
explicitly selected profiles of specific slices would lead to
similar results as if one would take functions related to more,
let's say, five but consecutive modes - from the fundamental to
the fourth order. However, adding field profiles from different
slices may lead to a (near) linear dependency of functions $X$,
and result in non-unique functions $Y$. Obviously, the safe choice
is to use in the approximation of a component of the total field
only profiles from a single slice. Nevertheless, when only a few
modal components are used, it may, as our calculations show, be
beneficial to use one or two modes from other slice(s).

Similar ideas can be applied to optical scattering problems in 2D
and 3D. A preliminary account of corresponding simulations has
been given in \cite{ISH08c}, \cite{ISK09}.

\section*{Acknowledgements}
This work was supported by the Dutch Technology Foundation (BSIK /
NanoNed project TOE.7143)

\bibliographystyle{unsrt}
\bibliography{wgcalc}

\end{document}